%% file: SubmissionCorr.tex
\documentclass[11pt]{article}
\RequirePackage{algorithm,algorithmic,amssymb,epsf,epic,graphicx}

\input{latexmac}

\denseformat

\begin{document}

\def\etal{{et al.~}}
\def\Sp{\mathit{Sp}}
\def\ReadDg{\mathit{Read\_Edge}}
\def\wmax{{{\hat \omega}}}
\def\ttl{\mathit{ttl}}
\def\Rnd{\mathit{Round}}
\def\deg{\mathit{deg}}
\def\ctr{\mathit{ctr}}
\def\ScDgs{\mathit{Scanned\_Edges}}
\def\SCD{\mathit{SCANNED}}
\def\NOTSCD{\mathit{NOTSCANNED}}
\def\CRASH{\mathit{CRASH}}
\def\CR{\mathit{CRASH}}
\def\CRASHED{\mathit{CRASHED}}
\def\SyncIncr{\mathit{Sync\_Incr}}
\def\actdeg{\mathit{actdeg}}
\def\OLD{\mathit{OLD}}
\def\NEW{\mathit{NEW}}
\def\mark{\mathit{mark}}
\def\AsyncRnd{\mathit{Async\_Rnd}}
\def\status{\mathit{status}}
\def\scstat{\mathit{scan\_status}}
\def\lab{\mathit{label}}
\def\seclabel{\mathit{sec\_label}}
\def\own{\mathit{own}}
\def\SELF{\mathit{SELF}}
\def\PEER{\mathit{PEER}}
\def\DynRnd{\mathit{Dyn\_Rnd}}
\def\Delete{\mathit{Delete}}
\def\XReplace{\mathit{XReplace}}
\def\UpdLab{\mathit{Update\_Label}}
\def\Crash{\mathit{Crash}}
\def\CrashLoop{\mathit{CrashLoop}}
\def\CrashItern{\mathit{CrashItern}}
\def\fetched{\mathit{fetched}}
\def\TRUE{\mathit{TRUE}}
\def\FALSE{\mathit{FALSE}}
\def\crashstat{\mathit{crash\_status}}
\def\wmax{\hat{\omega}}
\def\END{\mathit{END}}
\def\te{\tilde{e}}
\def\tu{\tilde{u}}
\def\che{\check{e}}
\def\chu{\check{u}}
\def\Old{\mathit{Old}}
\def\tR{\tilde{R}}
\def\eps{{\epsilon}}
\def\UpLab{\mathit{Update\_Label}}
\def\chP{\check{P}}
\def\VOID{\mathit{VOID}}

\newcommand {\ignore} [1] {}

\title{Balancing Degree, Diameter and Weight in Euclidean Spanners
\thanks
{A preliminary version of this paper appeared in ESA'10.}
}

\author{
Shay Solomon
\thanks{Department of Computer Science,
        Ben-Gurion University of the Negev, POB 653, Beer-Sheva 84105, Israel.
        \newline E-mail: {\tt \{shayso,elkinm\}@cs.bgu.ac.il}
     \newline Both authors are partially supported by the Lynn and William Frankel Center for Computer Sciences. 
   \newline {\mbox ~~~~}\textsection This research has been supported by the Clore Fellowship grant No.\ 81265410.}
      ~$^{\scriptsize \mbox \textsection}$   
         \and
Michael Elkin$^{\scriptsize \mbox \textdagger}$\thanks{This research has been supported by the BSF grant No.\ 2008430.}}

\date{\empty}

\begin{titlepage}
\def\thepage{}
\maketitle

\begin{abstract}
In a seminal STOC'95 paper, Arya et al.\ \cite{ADMSS95} devised a
construction that for any set $S$ of $n$ points in $\mathbb R^d$ and
any $\eps > 0$, provides a $(1+\eps)$-spanner with diameter $O(\log
n)$, weight $O(\log^2 n) \cdot w(MST(S))$, and constant maximum
degree. Another construction of \cite{ADMSS95} provides a
$(1+\eps)$-spanner with $O(n)$ edges and diameter $O(\alpha(n))$, where
$\alpha$ stands for the inverse Ackermann function. There are also a
few other known constructions of $(1+\eps)$-spanners. 
Das and Narasimhan
\cite{DN94} devised a construction with constant maximum degree and weight $O(w(MST(S)))$, but the
diameter may be arbitrarily large. In another construction by Arya et al.\ \cite{ADMSS95} there is diameter
$O(\log n)$ and weight $O(\log n) \cdot w(MST(S))$, but it may have
arbitrarily large maximum degree. While these constructions address
some important practical scenarios, they fail to address situations
in which we are prepared to compromise on one of the parameters, but
cannot afford this parameter to be arbitrarily large.

In this paper we devise a novel \emph{unified} construction that
trades between the maximum degree, diameter and weight gracefully. For a
positive integer $k$, our construction provides a
$(1+\eps)$-spanner with maximum degree $O(k)$, diameter $O(\log_k n
+ \alpha(k))$, weight $O(k \cdot \log_k n \cdot \log n) \cdot
w(MST(S))$, and $O(n)$ edges. Note that for $k = O(1)$ this gives
rise to maximum degree $O(1)$, diameter $O(\log n)$ and weight
$O(\log^2 n) \cdot w(MST(S))$, which is one of the aforementioned
results of \cite{ADMSS95}.  For $k= n^{1/\alpha(n)}$ this gives rise
to diameter $O(\alpha(n))$, weight $O(n^{1/\alpha(n)} \cdot \log n
\cdot \alpha(n)) \cdot w(MST(S))$ and maximum degree
$O(n^{1/\alpha(n)})$.
 In the  corresponding result from \cite{ADMSS95}
the spanner has the same number of edges and diameter, but its
weight and degree may be arbitrarily large.
 Our bound of $O(\log_k n + \alpha(k))$ on
the diameter is optimal under the constraints that the maximum
degree is $O(k)$ and the number of edges is $O(n)$. Similarly to the
bound of Arya et al.\ \cite{ADMSS95}, our bound on the weight is
optimal up to a factor of $\log n$.
Our construction also provides a similar tradeoff in the complementary range
of parameters, i.e., when the weight should be smaller than $\log^2 n$, but the diameter
is allowed to grow beyond $\log n$.

For random point sets in the $d$-dimensional unit cube, we ``shave'' a factor of $\log n$ 
from the weight bound. Specifically, in this case our construction achieves maximum
degree $O(k)$, diameter $O(\log_k n
+ \alpha(k))$ and weight that is with high probability $O(k \cdot \log_k n) \cdot
w(MST(S))$. 


Finally, en route to these results we devise optimal constructions of 1-spanners for general tree metrics, which are of independent interest. 
\end{abstract}
\end{titlepage}

\pagenumbering {arabic} 

\section{Introduction}
\subsection{Euclidean Spanners} 
Consider the weighted complete graph
$\mathcal S = (S,{S \choose 2})$ induced by a set $S$ of $n$ points
in $\mathbb R^d, d \ge 2$. The weight of an edge $(x,y) \in {S
\choose 2}$, for a pair of distinct points $x,y \in S$, is defined
to be the Euclidean distance $\|x-y\|$ between $x$ and $y$. Let $G = (S,E)$ 
be a spanning subgraph of
$\mathcal S$, with $E \subseteq {S \choose 2}$, and assume that exactly as in $\mathcal S$, for any
edge $e = (x,y) \in E$, its weight $w(e)$ in $G$ is defined to be
$\|x-y\|$. For a parameter $\eps > 0$, the spanning subgraph $G$ is
called a \emph{$(1+\eps)$-spanner} for the point set $S$ if for
every pair $x,y \in S$ of distinct points, the distance
$dist_G(x,y)$ between $x$ and $y$ in $G$ is at most $(1+\eps) \cdot
\|x-y\|$. Euclidean spanners were introduced\footnote{The notion
``spanner" was coined by Peleg and Ullman \cite{PU89}, who also
introduced spanners for general graphs.} 
in 1986
by Chew \cite{Chew86}. Since then they evolved into an important
subarea of Computational Geometry
\cite{Keil88,DJ89,RS91,KG92,ADDJS93,DHN93,DN94,ADMSS95,DNS95,AS97,RS98,AWY05,CG06,DES09tech}.
(See also the book by Narasimhan and Smid on Euclidean spanners \cite{NS07}, and the
references therein.) Also, Euclidean spanners have numerous
applications in geometric approximation algorithms
\cite{RS98,GLNS02,GLNS08}, geometric distance oracles
\cite{GLNS02,GNS05,GLNS08}, Network Design \cite{HP00,MP00} and in
other areas.

In many of these applications one is required to construct a
$(1+\eps)$-spanner $G = (S,E)$ that satisfies a number of useful
properties. First, the spanner should contain $O(n)$ (or nearly
$O(n)$) edges. Second, its {\em weight} $w(G) = \sum_{e
\in E}w(e)$ should not be much greater than the weight $w(MST(S))$ of
the minimum spanning tree $MST(S)$ of $S$. Third, its
\emph{diameter} $\Lambda = \Lambda(G)$ should be small,  i.e., for
every pair of points $x,y \in S$ there should exist a path $P$ in $G$
that contains at most $\Lambda$ edges and has weight $w(P) = \sum_{e
\in E(P)} w(e) \le (1+\eps) \cdot \|x-y\|$. Fourth, its {\em maximum
degree} (henceforth, {\em degree}) $\Delta(G)$ should be small.

In a seminal STOC'95 paper that culminated a long line of research,
Arya et al.\ \cite{ADMSS95} devised a construction of
$(1+\eps)$-spanners with lightness\footnote{For convenience, we
will henceforth refer to the normalized notion of weight $\Psi(G) = {{w(G)}
\over {w(MST(S))}}$, which we call {\em lightness}.} $O(\log^2 n)$, diameter $O(\log
n)$ and constant degree. They also devised a construction of
$(1+\eps)$-spanners with diameter $O(\alpha(n))$ (respectively, $O(1)$)
and $O(n)$ (resp., $O(n \cdot \log^* n)$) edges, where $\alpha$ stands for the inverse Ackermann function.
However,
in the latter construction the resulting spanners may have
arbitrarily large (i.e., $\Omega(n)$) lightness and degree. There are also a few other
known constructions of $(1+\eps)$-spanners. Das and Narasimhan
\cite{DN94} devised a construction with  constant  degree and
lightness, but the diameter may be arbitrarily large. 
(See also \cite{GLN02} for a faster implementation of a spanner construction with constant
degree and lightness.)
There is also
another construction by Arya et al.\ \cite{ADMSS95} that guarantees
that both the diameter and the lightness are $O(\log n)$,
 but the
degree may be arbitrarily large. While these constructions address
some important practical scenarios, they certainly do not address
all of them. In particular, they fail to address situations in which
we are prepared to compromise on one of the parameters, but cannot
afford this parameter to be arbitrarily large. 

In this paper we devise a novel \emph{unified} construction that
trades between the degree, diameter and weight gracefully. For a
positive integer $k$, our construction provides a
$(1+\eps)$-spanner with  degree $O(k)$, diameter $O(\log_k n +
\alpha(k))$, lightness $O(k \cdot \log_k n \cdot \log n)$,
 and $O(n)$ edges. Also, we can improve the bound on the
diameter from $O(\log_k n + \alpha(k))$ to $O(\log_k n)$, at the
expense of increasing the number of edges from $O(n)$ to $O(n \cdot
\log^* n)$. Note that for $k = O(1)$ our tradeoff gives rise to
degree $O(1)$, diameter $O(\log n)$ and lightness $O(\log^2 n)$,
which is one of the aforementioned results of \cite{ADMSS95}. Also, for
$k= n^{1/\alpha(n)}$ it gives rise to a spanner with degree
$O(n^{1/\alpha(n)})$,
 diameter $O(\alpha(n))$ and lightness $O(n^{1/\alpha(n)} \cdot \log n \cdot \alpha(n))$. In the corresponding result from \cite{ADMSS95} the spanner has the
same number of edges and diameter, but its lightness and degree may
be arbitrarily large.

In addition,  we can
achieve lightness $o(\log^2 n)$ at the expense of increasing the
diameter. Specifically, for a parameter $k$ the second variant of our construction
provides a
$(1+\eps)$-spanner with  degree $O(1)$, diameter $O(k \cdot
\log_k n)$, and lightness $O(\log_k n \cdot \log n)$.
For example, for $k = \log^{\delta} n$, for an arbitrarily small constant
 $\delta >0$, we get a
$(1+\eps)$-spanner with  degree $O(1)$, diameter $O(\log^{1+\delta} n)$ and lightness $O(\frac{\log^2 n}{\log \log
n})$.

Our unified construction can be implemented in $O(n \cdot \log n)$ time.
This matches the state-of-the-art
running time of the aforementioned constructions \cite{ADMSS95,GLN02}. 
See Table
\ref{tab1} for a concise comparison of previous and new
results.

\begin{table}
\begin{center}
\begin{tabular}{|c|c|c|c|c|c|c|c|c|}
\hline
 & \cite{ADMSS95}  & \cite{ADMSS95} & {\bf New} & {\bf New} &{\bf New}
&{\bf New} &{\bf New} &{\bf New}   \\
\hline
 & {\bf I,II} & & {\bf I} & {\bf I} & {\bf I} & {\bf II} & {\bf II}
& {\bf II} \\
\hline $k$ & $1$ & &  $\log^\delta n$ & $2^{\sqrt{\log n}}$ &
$n^{1/\alpha(n)}$ & $\log^\delta n$ & $2^{\sqrt{\log n}}$
& $n^\zeta$ \\
\hline \hline $\Delta$ & 1    & $n$ & $\log^\delta n$ &
$2^{\sqrt{\log n}}$ & $n^{1/\alpha(n)}$  & 1 & 1 & 1 \\
\hline $\Lambda$ & $\log n$   & $\alpha(n)$ &  ${{\log n} \over
{\log\log n}}$ & $\sqrt{\log n}$  & $\alpha(n)$  & $\log^{1+
\delta} n$ & $2^{O(\sqrt{\log
n})}$ & $n^\zeta$ \\
\hline $\Psi$    &  $\log^2 n$   & $n$ & $\log^{2 + \delta}
n$    & $2^{O(\sqrt{\log n})}$
& $n^{O(1/\alpha(n))}$ &${{ \log^2 n} \over {\log\log n}} $ & $\log^{3/2} n$ & $\log n$ \\
\hline
\end{tabular}
\end{center}
\caption[]{ \label{tab1} \footnotesize A concise comparison of previous and new
results. Each column corresponds to a set of parameters that can be
achieved simultaneously. For each column the first row indicates
whether the result is new or due to \cite{ADMSS95}. (The first
column is due to \cite{ADMSS95}, but can also be achieved from both
our tradeoffs.) For new results, the second row indicates whether it
is obtained by the first (I) or the second (II) tradeoff. (The first
tradeoff is degree $O(k)$, diameter $O(\log_k n + \alpha(k))$, and
lightness $O(k \cdot \log_k n \cdot \log n)$. The second tradeoff is
degree $O(1)$, diameter $O(k \cdot \log_k n)$ and lightness
$O(\log_k n \cdot \log n)$.) The third row indicates the value of
$k$ that is substituted in the corresponding tradeoff. The next
three rows indicate the resulting degree ($\Delta$), diameter
$(\Lambda)$ and lightness $(\Psi)$. The number of edges used in all
constructions is $O(n)$. To save space, the $O$ notation is omitted
everywhere except for the exponents. The letters $\delta$ and
$\zeta$ stand for arbitrarily small positive constants.
} 
\end{table}

\ignore{
[[S: This paragraph is very problematic]]
To summarize, our construction improves existing constructions for a number of specific sets of parameters.
For other sets of parameters it matches the current state-of-the-art bounds due to \cite{ADMSS95,GLN02}.
[[S: These two sentences are inaccruate; we only improve a single set of parameters (when the diameter
is $\alpha(n)$), and only match the current state-of the-art bound for another single set of parameters
(when the diameter is $O(\log n)$); moreover, we cannot obtain all state-of-the-art bounds known today
with our construction, for example, we cannot get constant degree and lightness like \cite{DN94,GLN02}.]]
We believe that replacing a number of separate ``ad-hoc'' constructions by a single unified parametereized
construction is an important step towards laying [[S:?]] more solid theoretical foundation to the area of Euclidean spanners.
[[S: This paragraph needs to be either removed or rewritten; the focus should be on the fact that we obtain
a continuous tradeoff rather than a few discrete sets of parameters, and this is an important step towards
a deep(er) understanding of Euclidean spanners]]
}

Note that in any construction of spanners with  degree $O(k)$, the
diameter is $\Omega(\log_k n)$. Also, Chan and Gupta \cite{CG06}
showed that any $(1+\eps)$-spanner with $O(n)$ edges must have
diameter $\Omega(\alpha(n))$. Consequently, our upper bound of
$O(\log_k n + \alpha(k))$ on the diameter is tight under the
constraints that the degree is $O(k)$ and the number of edges is
$O(n)$. If we allow $O(n \cdot \log^* n)$
 edges in the spanner, then our bound on the diameter is
reduced to $O(\log_k n)$, which is again tight under the constraint
that the degree is $O(k)$. 

In addition, Dinitz et al.\ \cite{DES09tech} showed that for any
construction of spanners, if the diameter is at most $O(\log_k n)$,
then the lightness  is at least $\Omega(k \cdot \log_k n) $ and vice
versa, if the lightness is at most $O(\log_k n) $, then the diameter is
at least $\Omega(k \cdot \log_k n)$. This lower bound implies that
the bound on lightness in both our tradeoffs  cannot possibly be
improved by more than a factor of $\log n$. The same slack of
$\log n$ is present in the result of \cite{ADMSS95} that
guarantees lightness $O(\log^2 n)$, diameter $O(\log n)$ and
constant degree.
\vspace{-0.05in}
\subsubsection{Euclidean Spanners for Random Point Sets} \label{sec111}
\vspace{-0.05in}
For random point sets in the $d$-dimensional unit cube (henceforth, unit cube),
we ``shave'' a factor of $\log n$ from the lightness bound in both
our tradeoffs,
and show that the first (respectively, second) variant of our construction achieves maximum
degree $O(k)$ (resp., $O(1)$), diameter $O(\log_k n
+ \alpha(k)$) (resp., $O(k \cdot \log_k n)$) and lightness that is with high probability (henceforth, w.h.p.) $O(k \cdot \log_k n)$ (resp., $O(\log_k n)$). 
Note that for $k=O(1)$ both these tradeoffs give rise to degree $O(1)$, diameter $O(\log n)$ and lightness (w.h.p.) $O(\log n)$.
In addition to these tradeoffs, we can get a construction of $(1+\eps)$-spanners with diameter $O(\log n)$ and lightness (w.h.p.) $O(1)$. 

\vspace{-0.05in}
 \subsubsection{Spanners for Doubling Metrics} \label{sec112}
 \vspace{-0.05in}
The \emph{doubling dimension} of a metric $(X,\delta)$ is the smallest value $\zeta$
such that every ball $B$ in the metric can be covered by at most
$2^\zeta$ balls of half the radius of $B$. The metric $(X,\delta)$ is called \emph{doubling}
if its doubling dimension $\zeta$ is constant. 
Spanners for doubling metrics have received much attention in recent years (see, e.g., \cite{CGMZ05,HPM05,CG06,GR08}).
In particular, Chan et al.\  \cite{CGMZ05} showed that for any doubling metric $(X,\delta)$ there exists a $(1+\eps)$-spanner
with constant maximum degree, but this spanner may have arbitrarily large diameter. In addition, Chan and Gupta \cite{CG06} devised a construction of $(1+\eps)$-spanners for doubling metrics
that achieves the optimal tradeoff between the number of edges and diameter, but these spanners may have arbitrarily large degree. We present a single construction of $O(1)$-spanners
for doubling metrics
that achieves the optimal tradeoff between the degree, diameter and number of edges
in the entire range of parameters.
Specifically, for a parameter $k$, our construction provides an $O(1)$-spanner with maximum degree $O(k)$, diameter $O(\log_k n + \alpha(k))$,
and $O(n)$ edges.
Also, we can improve the bound on the diameter from $O(\log_k n + \alpha(k))$ to $O(\log_k n)$, at the
expense of increasing the number of edges from $O(n)$ to $O(n \cdot
\log^* n)$. More generally, we can achieve the same optimal tradeoff between the number of edges and diameter
as the spanners of \cite{CG06} do, while also having the optimal maximum degree.
The drawback is, however, that the stretch of our
spanners is $O(1)$ rather than $1+\eps$.

\subsection{Spanners for Tree Metrics}
Let $\vartheta_n$ be the metric induced by $n$ points $v_1,v_2,\ldots,v_n$ lying on
the $x$-axis with coordinates $1,2,\ldots,n$, respectively.
 In a classical STOC'82 paper
\cite{Yao82}, Yao showed that there exists a 1-spanner\footnote{The
graph $G$ is said to be a \emph{1-spanner} for $\vartheta_n$ if for
every pair of distinct vertices $v_i,v_j \in V$, the distance
between them in $G$ is equal to their distance $\|i-j\|$ in
$\vartheta_n$. Yao stated this problem in terms of
partial sums. However, the two statements of the problem are equivalent.} 
$G = (V,E)$ for $\vartheta_n$ with 
diameter $O(\alpha(n))$ and $O(n)$ edges, and that this is
tight. Chazelle \cite{Chaz87} extended the result of \cite{Yao82} to arbitrary
tree metrics. Other proofs of Chazelle's result appeared in
\cite{AS87,BTS94,Thor97,Sol11}. Thorup
\cite{Thor97} also devised an efficient parallel algorithm for
computing this 1-spanner. The problem was also studied for planar
metrics \cite{Thor95}, general metrics \cite{Thor92} and even for general graphs \cite{BGJRW09}. (See
also Chapter 12 in \cite{NS07} for an excellent survey on this
problem.)
The problem is also closely related to the extremely well-studied problem of
computing partial-sums. (See the papers of Tarjan \cite{Tarj79}, Yao \cite{Yao82},
Chazelle and Rosenberg \cite{CR91}, ~P\u{a}tra\c{s}cu and Demaine
\cite{PD04}, and the references therein.) For a discussion about the
relationship between these two problems see the introduction
of \cite{AWY05}. 

In all constructions \cite{Yao82,Chaz87,AS87,BTS94,Thor97,Sol11} of
1-spanners for tree metrics, the degree and lightness of the
resulting spanner may be arbitrarily large. Moreover, the constraint
that the diameter is $O(\alpha(n))$ implies that the
degree must be $n^{\Omega(1/\alpha(n))}$. A similar lower bound on
the lightness follows from the result of \cite{DES09tech}.

En route to our tradeoffs for Euclidean spanners, we have extended
the results of \cite{Yao82,Chaz87,AS87,BTS94,Thor97,Sol11} and devised a
construction that achieves the \emph{optimal} (up to constant
factors) \emph{tradeoff between all involved parameters}.
Specifically, consider an $n$-vertex tree $T$ of degree $\Delta(T)$,
and let $k$ be a positive integer. Our construction provides a
1-spanner for the tree metric $M_T$ induced by $T$ with 
degree $O(\Delta(T)+k)$, diameter $O(\log_k n + \alpha(k))$, 
lightness $O(k \cdot \log_k n)$, and $O(n)$ edges. We can also get a spanner with diameter $O(\log_k n)$, 
$O(n
\cdot \log^* n)$ edges, 
and the same degree
and lightness as above. For the complementary range of diameter,
the second variant of our construction provides a $1$-spanner with
degree $O(\Delta(T))$, diameter $O(k \cdot \log_k n)$,
lightness $O(\log_k n)$, and $O(n)$ edges. As was mentioned above, both these tradeoffs
are \emph{optimal up to constant factors}. 

We show that this general tradeoff between various parameters of
1-spanners for tree metrics is useful for deriving new results (and
improving existing results) in the context of Euclidean spanners and spanners
for doubling metrics. We
anticipate that this tradeoff would be found useful in the context of partial sums problems, and
for other
applications. 
 
\subsection{Our and Previous Techniques} \label{sec13}
The starting point for our construction is the construction of Arya
et al.\ \cite{ADMSS95} that achieves diameter $O(\log n)$, lightness
$O(\log^2 n)$ and constant degree. The construction of
\cite{ADMSS95} is built in two stages. First, a construction for the
1-dimensional case is devised. Then the 1-dimensional construction
is extended to arbitrary constant dimension. For 1-dimensional
spaces Arya et al.\ \cite{ADMSS95} start with devising a
construction of 1-spanners with diameter, lightness and degree all
bounded by $O(\log n)$. This construction is quite simple; it is
essentially a flattened version of a deterministic skip-list. Next,
by a more involved argument they show that the degree can be reduced
to $O(1)$, at the expense of increasing the stretch parameter from 1
to $1+\eps$. Finally, the generalization of their construction to
point sets in the plane (or, more generally, to $\mathbb R^d$) is
far more involved. Specifically, to this end Arya et al.\
\cite{ADMSS95} employed two main tools. The first one is the
\emph{dumbbell trees}, the theory of which was developed by Arya et
al.\ in the same paper \cite{ADMSS95}. (See also Chapter 11 of
\cite{NS07}.) The second one is the bottom-up clustering technique
that was developed by Frederickson \cite{Fred93} for topology trees.
Roughly speaking, the \emph{Dumbbell Theorem} of \cite{ADMSS95}
states that for every point set $S$, one can construct a forest
$\mathcal D$ of $O(1)$ dumbbell trees, in which there
exists a tree $T \in \mathcal D$ for every pair $x,y$
of points from $S$, such that the distance $dist_T(x,y)$ between
$x$ and $y$ in $T$ is at most $(1+\eps)$ times their
Euclidean distance $\|x-y\|$.  
Arya et al.\
employ Frederickson's clustering technique on each of these $O(1)$
dumbbell trees to obtain their ultimate spanner.

Similarly to \cite{ADMSS95}, we start with devising a construction
of 1-spanners for the 1-dimensional case. However, our construction
achieves both diameter and lightness at most $O(\log n)$, in
conjunction with the \emph{optimal degree} of at most 3.\footnote{Observe that any graph (not necessarily 1-spanner) 
with maximum degree 2 must have diameter at least $\frac{n-1}{2}$.}
(Note that \cite{ADMSS95} paid
for decreasing the degree from $O(\log n)$ to $O(1)$ by increasing
the stretch of the spanner from 1 to $1+\eps$. Our construction
achieves stretch 1 in conjunction with logarithmic diameter and
lightness, and with the optimal degree.) Moreover, our construction is
far more general, as it provides the \emph{entire suite} of all
possible values of diameter, lightness and degree, and it is
\emph{optimal up to constant factors} in the entire range of
parameters. We then proceed to extending it to arbitrary tree
metrics. Finally, we employ the dumbbell trees of Arya et al.\
\cite{ADMSS95}. Specifically, we construct our 1-spanners for the
metrics induced by each of these dumbbell trees, and return their
union as our ultimate spanner. As a result we obtain a
\emph{unified} construction of Euclidean spanners that achieves
near-optimal tradeoffs in the entire range of
parameters. We remark that it is unclear whether the construction of
Arya et al.\ \cite{ADMSS95} can be extended to provide additional
combinations between the diameter and lightness other than $O(\log n)$
and $O(\log^2 n)$, respectively; roughly speaking, the logarithms
there come from the number of levels in Frederickson's topology
trees. In particular, the construction of Arya et al.\
\cite{ADMSS95} that achieves diameter $O(\alpha(n))$ and arbitrarily
large lightness and degree is based on completely different ideas.
On the other hand, our construction yields a stronger result
(diameter $O(\alpha(n))$, lightness and degree
$n^{O(1/\alpha(n))}$), and this result is obtained by substituting a
different parameter into one of our tradeoffs. Moreover, our construction is much simpler and 
more modular than that of \cite{ADMSS95}. In particular, it
does not employ Frederickson's bottom-up clustering technique, but
rather constructs 1-spanners for dumbbell trees directly. 

Also, our construction of 
1-spanners for tree metrics (that we use for dumbbell trees)
is fundamentally different from the previous
constructions due to \cite{Yao82,Chaz87,AS87,BTS94,Thor97,Sol11}. 
In particular, the techniques of \cite{Chaz87,AS87,BTS94,Thor97,Sol11} 
for generalizing constructions of 1-spanners from 1-dimensional 
metrics to general tree metrics
ensure that the diameter of the resulting
spanners is not (much) greater than the diameter in
the 1-dimensional case. However,
 the degree and/or lightness of spanners for tree metrics
that are obtained by these techniques may be arbitrarily large.  To
overcome this obstacle we adapt the techniques of
\cite{Chaz87,AS87,BTS94,Sol11} to our purposes. 
Next, we overview this adaptation.
A central ingredient in the generalization techniques of
\cite{Chaz87,AS87,BTS94,Sol11} is a tree decomposition procedure. Given an
$n$-vertex rooted tree $(T,rt)$ and a parameter $k$, this procedure
computes a set $C$ of
$O(k)$ \emph{cut vertices}. This set has the property that removing all
vertices of $C$ from the tree $T$ decomposes $T$ into a collection
$\mathcal F$ of trees, so that each tree $\tau \in \mathcal F$
contains $O(n/k)$ vertices. This decomposition induces a tree
$\mathcal Q = \mathcal Q(\tau,C)$  over the vertex set $C \cup \{rt\}$ in a natural
way: a cut vertex $w \in C$ is defined to be a child of its closest
ancestor in $T$ that belongs to $C \cup \{rt\}$. For our purposes, it is crucial that 
the degree of the tree $\mathcal Q$ will not be (much) greater than the degree 
of $T$. In addition, it is essential that
each tree $\tau \in \mathcal F$ will be incident to at most $O(1)$
cut vertices. We devise a novel decomposition procedure that
guarantees these two basic properties. Intuitively, our decomposition
procedure ``slices" the tree in a ``path-like" fashion. This
path-like nature of our decomposition enables us to keep the degree
and lightness of our construction for general tree metrics (essentially) as small as in the 1-dimensional case.

\subsection{Structure of the Paper}
In Section \ref{sec2} we describe our construction of 1-spanners for tree metrics. Therein
we start (Section \ref{sec21}) with outlining our basic scheme.  We proceed (Section \ref{onedim}) with describing our 1-dimensional construction.
In Section \ref{arbtrees} we extend this construction to general tree metrics. 
Our tree decomposition procedure (which is in the heart of this extension) is described in Section \ref{sec231}.
In Section \ref{imp} we derive our results for Euclidean spanners and spanners for doubling metrics.
 
\subsection{Preliminaries} \label{prel}
An \emph{$n$-point metric space} $M = (V,dist)$ can be viewed as the
complete graph $G(M) = (V,{V \choose 2},dist)$ in which for every
pair of points $x,y \in V$, the weight of the edge $e = (x,y)$ in
$G(M)$ is defined by $w(x,y) = dist(x,y)$. Let $G$ be a spanning
subgraph of $M$. 
We say that $G$ is a \emph{$t$-spanner} for $M$
if for every pair $x,y \in V$ of distinct points, there exists a path in $G$
between $x$ and $y$ whose weight (i.e., the sum of all edge weights in it)
is at most $t \cdot dist(x,y)$. Such a path is called a \emph{$t$-spanner path}.
The \emph{stretch} of $G$ is the minimum number $t$, such that $G$ is a $t$-spanner
for $M$.
Let $T$ be an arbitrary tree, and denote by $V(T)$ the vertex set of $T$.
For any two vertices $u,v$ in $T$, their (weighted) distance
in $T$ is denoted by $dist_T(u,v)$. The tree metric $M_T$ induced by
$T$ is defined as $M_T = (V(T),dist_T)$. 
The \emph{size} of $T$,
denoted $|T|$, is the number of vertices in $T$.
Finally, for a positive integer $n$, we denote the set $\{1,2,\ldots,n\}$ by
$[n]$. 

\section{1-Spanners for Tree Metrics} \label{sec2}
 
\subsection{The Basic Scheme} \label{sec21}
Consider an arbitrary $n$-vertex (weighted) rooted tree $(T,rt)$, and let $M_T$ be
the tree metric induced by $T$. Clearly, $T$ is both a 1-spanner and
an MST of $M_T$, but its diameter may be arbitrarily large. We would like to reduce the diameter
of this 1-spanner by adding to it some edges. On the other hand, the number of edges of the resulting spanner should still be linear in $n$. Moreover, the lightness and the maximum degree of the resulting
spanner should also be reasonably small.

Let $H$ be a spanning subgraph of $M_T$. The \emph{monotone distance} between any two
points $u$ and $v$ in $H$
is defined as the minimum number of edges in a 1-spanner path in $H$
connecting them.
Two points in $M_T$ are called \emph{comparable} if one is an ancestor of the other
in the underlying tree $T$.
The \emph{monotone diameter} (respectively, \emph{comparable monotone diameter}) of $H$, denoted $\Lambda(H)$ (resp., $\bar \Lambda(H)$), is defined as
the maximum monotone distance in $H$ between any two points (resp., any two comparable points) in $M_T$.
Observe that if any two \emph{comparable} points are
connected via a 1-spanner path that consists of at most $h$ edges,
then any two \emph{arbitrary} points are connected via a 1-spanner path that
consists of at most
$2h$ edges. Consequently, $\bar \Lambda(H) \le \Lambda(H) \le 2\cdot \bar \Lambda(H)$. We henceforth restrict the attention to
comparable monotone diameter in the sequel.

Let $k\ge 2$ be a fixed parameter. The first ingredient of the algorithm is to
select a set of $O(k)$ \emph{cut vertices} whose removal from
$T$ partitions it into a collection of subtrees of
size $O(n/k)$ each. (As mentioned in the last paragraph of Section \ref{sec13}, we also require this set to satisfy several additional properties.)  Having selected the cut vertices,
the next step of the algorithm is to connect the cut vertices via
$O(k)$ edges,
so that the monotone distance between any pair of comparable cut vertices will be 
small. (This phase does not involve a recursive call of the
algorithm.)
Finally, the algorithm calls itself recursively for each of the
subtrees.

We insert all edges of the original tree $T$ into our final spanner $H$. These edges connect between cut vertices and subtrees
in the spanner. We remark that the spanner contains no other edges that connect between cut vertices and subtrees.
Moreover, the spanner contains no edges that connect between 
different subtrees. 
\subsection{1-Dimensional Spaces} \label{onedim}
In this section we devise an optimal construction of 1-spanners for $\vartheta_n$. 
(Recall that $\vartheta_n$ is the metric induced by $n$ points $v_1,v_2,\ldots,v_n$
lying on the $x$-axis with coordinates $1,2,\ldots,n$, respectively.)
Our argument extends easily to any 1-dimensional space.

Denote by $P_n$ the path $(v_1,v_2),(v_2,v_3),\ldots,(v_{n-1},v_n)$ that induces the metric $\vartheta_n$.
We remark that the edges of $P_n$ (henceforth, \emph{path-edges}) belong to all spanners that we construct.
\subsubsection{Selecting the Cut-Vertices} \label{sec221}
\begin{figure*}[htp]
\begin{center}
\begin{minipage}{\textwidth} 
\begin{center}
\setlength{\epsfxsize}{4in} \epsfbox{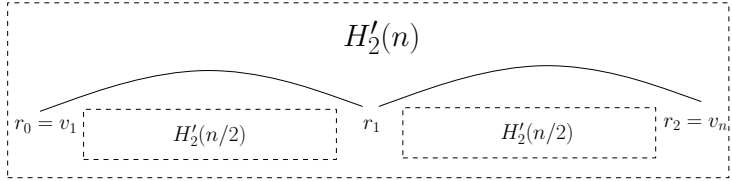}
\end{center}
\end{minipage}
\caption[]{ \label{dim} \sf \footnotesize The construction for $k=2$. Only
the first level of the recursion is illustrated. (Path-edges are not depicted in the figure.) The
cut vertex $r_1 = v_{n/2}$ is connected via edges to the two
sentinels $v_1$ and $v_n$. The construction proceeds recursively for
each of the two intervals $I_1$ and $I_2$.}
\end{center}
\end{figure*}

Let $k \ge 2$ be a fixed parameter.
The task of selecting the cut vertices in the 1-dimensional case is
trivial. (We assume for simplicity that $n$ is an integer power of
$k$.) In addition to the two endpoints $v_1$ and $v_n$ of the path,
we select the $k-1$ points $r_1, r_2,\ldots,r_{k-1}$ to be
cut vertices, where for each $i \in [k-1]$, $r_i = v_{i(n/k)}$.
Indeed, by removing the $k+1$ cut vertices $r_0 =
v_1,r_1,\ldots,r_{k-1},r_k=v_n$ from the path (along with their
incident edges), we are left with $k$ intervals $I_1,I_2,\ldots,I_k$ of length at most
$n/k$ each. The two endpoints $v_1$ and $v_n$ of the path are called
the \emph{sentinels}, and they play a special role in the
construction. (See Figure \ref{dim} for an illustration for the case
$k=2$.)

\subsubsection{1-Spanners with Low Diameter} \label{sec222}
In this section we devise
a construction $H_k(n)$ of 1-spanners for $\vartheta_n$ with
comparable monotone diameter $\bar \Lambda(n) = \bar
\Lambda(H_k(n))$ in the range $\Omega(\alpha(n)) = \bar \Lambda(n) =
O(\log n)$. In Section \ref{sec223} we turn our attention to spanners with larger
monotone diameter.

\begin{figure*}[htp]
\begin{center}
\begin{minipage}{\textwidth} 
\begin{center}
\setlength{\epsfxsize}{4.4in} \epsfbox{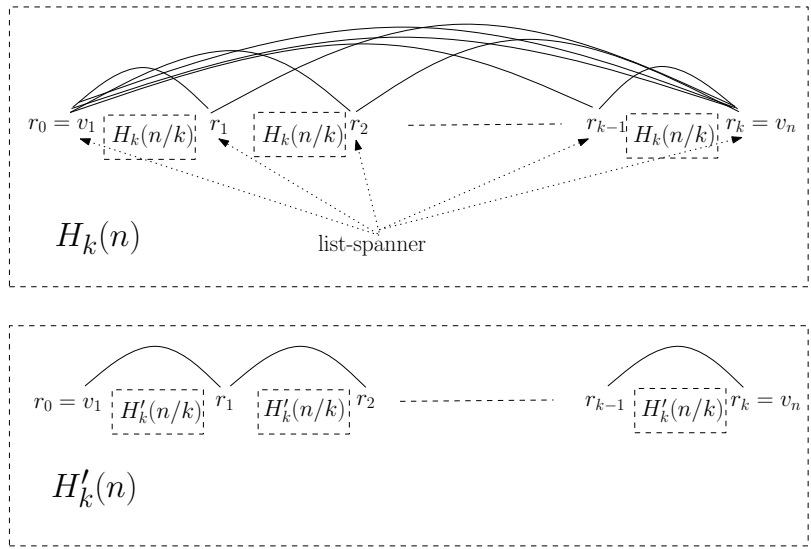}
\end{center}
\end{minipage}
\caption[]{ \label{dim2} \sf \footnotesize The constructions
$H_k(n)$ and $H'_k(n)$ for a general parameter $k, k \ge 2$. Only the first level of
the recursion is illustrated. (Path-edges are not depicted in the
figure.) For $H_k(n)$, all
the cut vertices are connected via the list-spanner, and, in
addition, each of the two sentinels is connected to all other $k$
cut vertices. For $H'_k(n)$, each cut vertex $r_{i-1}$ is connected to
the next cut vertex $r_{i}$ in line, $i \in [k]$.}
\end{center}
\end{figure*}

First, the algorithm connects the $k+1$ cut vertices
$r_0=v_1,r_1,\ldots,r_{k-1},r_{k} = v_n$ via one of the
aforementioned constructions of 1-spanners from
\cite{Yao82,Chaz87,AS87,BTS94,Thor97,Sol11} (henceforth,
\emph{list-spanner}). In other words, $O(k)$ edges between
cut vertices are added to the spanner $H_k(n)$ to guarantee that the monotone distance in the spanner between any two
cut vertices\footnote{In the 1-dimensional case any two points are
comparable.} will be $O(\alpha(k))$.
%
Then the algorithm adds to the spanner $H_k(n)$ edges
that connect each of the two sentinels to all other $k$ cut vertices.
Finally, the algorithm calls itself recursively for each of the
intervals $I_1,I_2,\ldots,I_k$. At the bottom level of the recursion, i.e., when $n \le k$,
the algorithm uses the list-spanner to connect all points, and, in
addition, it adds to the spanner edges that connect each of the two sentinels $v_1$ and $v_n$ to all the
other $n-1$ points. (See Figure \ref{dim2} for an illustration.)

Denote by $E(n)$ the number of edges in $H_k(n)$, excluding edges of
$P_n$. Clearly, $E(n)$ satisfies the recurrence $E(n) \le O(k) + k
\cdot E(n/k)$, with the base condition $E(q) = O(q)$, for all $q \le k$,
yielding $E(n) = O(n)$. Denote by $\Delta(n)$ the maximum degree of
a vertex in $H_k(n)$, excluding edges of $P_n$. Clearly, $\Delta(n)$
satisfies the recurrence $\Delta(n) \le \max\{k,\Delta(n/k)\}$, with
the base condition $\Delta(q) \le q-1$, for all $q \le k$, yielding
$\Delta(n) \le k$. Including edges of $P_n$, the number of edges
increases by $n-1$ units, and the maximum degree increases by at
most two units.

Denote by $w(n)$ the weight of $H_k(n)$, excluding edges of $P_n$. 
Note that at most $O(k)$ edges are added
between cut vertices. Each of these edges has weight at
most $n-1$.
The total weight of all edges within an interval $I_i$ is at most
$w(n/k)$. 
Hence $w(n)$
satisfies the recurrence $w(n) \le O(n \cdot k) + k \cdot w(n/k)$, with the base condition
$w(q) = O(q^2)$, for all $q \le k$. It follows that $w(n) = O(n \cdot k \cdot \log_k n) = O(k
\cdot \log_k n) \cdot w(MST(\vartheta_n))$.
Including edges of $P_n$, the weight increases by $w(P_n) = n-1$ units.

Next, we show that the comparable monotone diameter $\bar \Lambda(n)$ of $H_k(n)$ is at most $O(\log_k n + \alpha(k))$.
The \emph{monotone radius} $R(n)$ of $H_k(n)$ is defined as
the maximum monotone distance in $H_k(n)$ between one of the sentinels (either $v_1$
or $v_n$) and some other point in $\vartheta_n$. Let $v_j$ be a point in $\vartheta_n$, and let $i$
be the index such that $v_j \in \{r_i\} \cup I_i$. (In other words, $i$ is the index such that
$i  (n/k) \le j < (i+1)  (n/k)$.) 
If $j = i  (n/k)$ then $v_j$ is the cut vertex $r_i$; in this case the $1$-spanner path $\Pi = \Pi(v_1,v_j)$ in $H_k(n)$
connecting the sentinel $v_1$ and the point $v_j$ will consist of the single edge $(v_1,v_j)$.
Otherwise, $j > i  (n/k)$ and $v_j \in I_i$. In this case the path $\Pi$ 
will start with the two edges $(v_1,v_{i(n/k)})$,
$(v_{i(n/k)},v_{i(n/k)+1})$. The point $v_{i(n/k)+1}$ is a sentinel of the $i$th interval $I_i$. 
Hence, the path $\Pi$ will continue recursively, from $v_{i(n/k)+1}$ to $v_j$. It follows that the monotone radius $R(n)$
satisfies the recurrence $R(n) \le 2+ R(n/k)$, with the base condition $R(q) = 1$, for all $q \le k$, yielding
$R(n) = O(\log_k n)$.
\ignore{
[[S: SSSSSSSSSSSS Next, we bound the comparable monotone diameter $\bar \Lambda(n)$ of $H_k(n)$.
Consider two arbitrary points $v_i,v_j$ in $\vartheta_n$, $i < j$,
and let $P = (v_i,v_{i+1},\ldots,v_j)$ be the sub-path of $P_n$ between $v_i$ and $v_j$.
We distinguish between three cases.
\\\emph{Case 1: $P$ contains no cut vertices.} In this case,
$v_i$ and $v_j$ belong to the same interval $I_\ell$, for some index $\ell \in [k-1]$, and then 
it can be argued inductively that their monotone distance is at most 
$\bar \Lambda(n/k)$.
\\\emph{Case 2: $P$ contains one cut vertex, denoted $r_\ell$.}
Otherwise, denote by $r$ and $r'$ the first and last cut vertices on $P$.
(It is possible that $r = v_i$ and $r' = v_j$.)
If $v_i$ and $v_j$ are cut vertices, then they 
By construction, $r$ and $r'$ are 
are connected (via the list-spanner) by a 1-spanner path
that consists of at most $O(\alpha(k))$ edges.
their monotone distance is
]]
}
It is easy to verify  that $\bar \Lambda(n)$ satisfies the
recurrence $\bar \Lambda(n) \le \max\{\bar \Lambda(n/k),O(\alpha(k)) + 2R(n/k) \}$, 
with the base condition $\bar \Lambda(q)
= O(\alpha(q))$, for all $q \le k$. Hence $\bar \Lambda(n) = O(\log_k n + \alpha(k))$. 

Denote the worst-case running time of the algorithm by $t(n)$, excluding the time needed to
add the edges of $P_n$ to the spanner.
We remark that the {list-spanner} of \cite{Yao82,Chaz87,AS87,BTS94,Thor97,Sol11} can
be implemented in linear time.
By construction,
$t(n)$ satisfies the recurrence $t(n) \le O(k) + k
\cdot t(n/k)$, with the base condition $t(q) = O(q)$, for all $q \le k$,
yielding $t(n) = O(n)$. 
Hence, the overall running time of the algorithm is $O(n)$.

Finally, we remark that the maximum degree of this construction can be easily reduced from $k+2$ to $k+1$, 
without increasing any of the other parameters by more than a constant factor;
the details of this technical argument are omitted.
In particular, for $k=2$ we will get this way the optimal degree 3, together with diameter and lightness $O(\log n)$;
the same result also follows from Theorem \ref{cor2} below.

\begin{theorem} \label{cor1} For any
$n$-point 1-dimensional space and a parameter $k \ge 2$,
there exists a 1-spanner with maximum degree at most
$k+1$, diameter $O(\log_k n + \alpha(k))$, lightness $O(k \cdot
\log_k n)$, and $O(n)$ edges.
The running time of this construction is $O(n)$.
\end{theorem}

\subsubsection{1-Spanners with High Diameter} \label{sec223}
In this section we devise a construction $H'_k(n)$ of 1-spanners for
$\vartheta_n$ with comparable monotone diameter $\bar \Lambda'(n) = \bar \Lambda(H'_k(n))$ in the range $\bar \Lambda'(n) = \Omega(\log n)$.

The algorithm connects the $k+1$ cut vertices
$r_0=v_1,r_1,\ldots,r_{k-1},r_k = v_n$ via a path of length
$k$, i.e., it adds the edges $(r_0,r_1),(r_1,r_2),\ldots,(r_{k-1},r_k)$
into the spanner. In addition, it calls itself recursively for each of the intervals $I_1,I_2,\ldots,I_k$. At
the bottom level of the recursion, i.e., when $n \le k$,
the algorithm adds no additional edges to the spanner.
(See Figures \ref{dim} and \ref{dim2} for
an illustration.)

Denote by $\Delta'(n)$ the maximum degree of a vertex in
$H'_k(n)$, excluding edges of $P_n$. Clearly, $\Delta'(n)$ satisfies
the recurrence $\Delta'(n) \le \max\{2,\Delta'(n/k)\}$, with the
base condition $\Delta'(q) = 0$, for all $q \le k$, yielding $\Delta'(n)
\le 2$. Including edges of $P_n$,
the maximum degree increases by at most two units, and so
$\Delta(H'_k(n)) \le 4$. Consequently, the number of edges in
$H'_k(n)$ is no greater than $2n$.

Denote by $w'(n)$ the weight of $H'_k(n)$, excluding edges of $P_n$.
Note that the weight of the
path connecting  
all $k+1$ cut vertices is equal to $n-1$. The total weight of all edges within an interval $I_i$
is at most $w'(n/k)$.
Hence $w'(n)$ satisfies the recurrence $w'(n) \le n-1 + k \cdot w'(n/k)$, with the base condition $w'(q)
\le q-1$, for all $q \le k$. It follows that $w'(n) = O(n \cdot \log_k n) = O(\log_k n) \cdot
w(MST(\vartheta_n))$. Including edges of $P_n$, the weight increases by $w(P_n) = n-1$ units.

Note that the monotone radius $R'(n)$ of $H'_k(n)$ satisfies the
recurrence $R'(n) \le k+ R'(n/k)$, with the base condition $R'(q)
\le q-1$, for all $q \le k$. Hence, $R'(n) = O(k \cdot \log_k n)$. 
Using reasoning similar to that of Section \ref{sec222}, 
we get that the comparable monotone diameter $\bar \Lambda'(n) =
\bar \Lambda(H'_k(n))$ of $H'_k(n)$ satisfies the
recurrence $\bar \Lambda'(n) \le \max\{\bar \Lambda'(n/k),k+
2R'(n/k)\}$, with the base condition $\bar \Lambda'(q) \le q-1$, for
all $q \le k$. It follows that $\bar \Lambda'(n) = O(k \cdot \log_k n)$.

We remark that the spanner $H'_k(n)$ is a planar graph.

Denote the worst-case running time of the algorithm by $t'(n)$, excluding the time needed to
add the edges of $P_n$ to the spanner.
It is easy to see that $t'(n)$ satisfies the recurrence $t'(n) \le O(k) + k
\cdot t'(n/k)$, with the base condition $t'(q) = O(1)$, for all $q \le k$,
yielding $t'(n) = O(n)$. 
Hence, the overall running time of the algorithm is $O(n)$.

Finally, similarly to the construction of Section \ref{sec222}, the maximum degree of this construction can be reduced from $4$ to $3$, without increasing
any of the other parameters by more than a constant factor. 
\begin{theorem} \label{cor2}
For any $n$-point 1-dimensional space and a parameter $k$, there exists a 1-spanner with maximum degree 3,
diameter $O(k \cdot \log_k n)$, and lightness $O(\log_k n)$.
Moreover, this 1-spanner is a planar graph.
The running time of this construction is $O(n)$.
\end{theorem}

\subsection{General Tree Metrics} \label{arbtrees}
In this section we extend the constructions of Section \ref{onedim} from line metrics
to general tree metrics.
\subsubsection{Selecting the Cut-Vertices} \label{sec231}
In this section we present a procedure for selecting,
given a tree $T$, a subset of $O(k)$ vertices whose removal
from the tree partitions it into
subtrees of size $O(|T|/k)$ each. This subset will also satisfy several
additional useful properties.

Let $(T,rt)$ be a rooted tree. 
For an inner
vertex $v$ in $T$ with $ch(v)$ children, we denote its children, from left to right, 
by
$c_1(v),c_2(v),\ldots,c_{ch(v)}(v)$.
Suppose without loss of generality
that the size of the subtree $T_{c_1(v)}$ of $v$ is no
smaller than the size of any other subtree of $v$, i.e.,
$|T_{c_1(v)}| \ge |T_{c_2(v)}|, |T_{c_3(v)}|, \ldots,
|T_{c_{ch(v)}(v)}|$. 
(This assumption can be guaranteed by a
straightforward procedure that runs in linear time.) 
We say that the vertex $c_1(v)$ 
is the \emph{left-most} child 
of $v$.
Also, 
an edge in $T$ is called \emph{left-most} if it connects a vertex
$v$ in $T$ and its left-most child $c_1(v)$.
We denote by $P(v) =
(v,c_1(v),\ldots,l(v))$ the path of left-most edges leading down
from $v$ to some leaf $l(v)$ in the subtree $T_v$ of $T$ rooted at $v$; 
the leaf $l(v)$ is referred to as
the \emph{left-most vertex} in $T_v$.
Also, let $l(T) = l(rt)$ denote the left-most vertex in the entire tree $T$. 
An inner vertex $v$ in $T$ is called \emph{$d$-balanced}, for $d \ge 1$, 
or simply
\emph{balanced} if $d$ is clear from the context, if $|T_{c_1(v)}|
\le |T|-d$. The first (i.e., closest to $v$) balanced vertex along $P(v)$ is denoted by
$b(v)$; if no vertex along $P(v)$ is balanced, we write $b(v) = NULL$.
Observe that for $|T| \ge 2d$, we have $|T|-d \ge d \ge 1$; in this case 
the one-before-last vertex along $P(v)$ (namely, the parent $\pi(l(v))$ of $l(v)$ in $T$) is balanced. 
Hence, in this case $b(v) \ne NULL$.

\begin{figure*}[htp]
\begin{center}
\begin{minipage}{\textwidth} 
\begin{center}
\setlength{\epsfxsize}{3.6in} \epsfbox{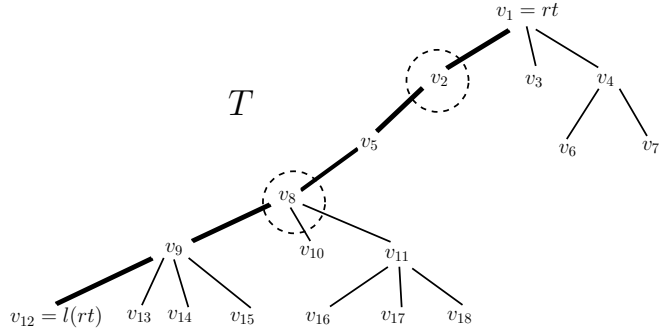}
\end{center}
\end{minipage}
\caption[]{ \label{arbi} \sf \footnotesize A rooted tree $(T,rt)$ with $n=|T|=18$
vertices $v_1=rt,v_2,\ldots,v_{18}$. The edges of $P(rt)$ are depicted by bold lines.
The first $6$-balanced vertex along $P(rt)$ is $v_2$. 
The procedure $CV$ on input $(T,rt)$ and $d=6$ returns the subset
$\{v_2,v_8\}$.}
\end{center}
\end{figure*}

Next, we present the Procedure $CV$ (standing for cut vertices) that accepts as input
a rooted tree $(T,rt)$ and a parameter $d \ge 1$, and returns as output a subset of $V(T)$.
If $|T| < 2d$, the
procedure returns the empty set $\emptyset$. Otherwise $|T| \ge 2d$, 
and so
the first balanced vertex $b = b(rt)$ along $P(rt)$
satisfies $b \ne NULL$.
In this case for each child
$c_i(b)$ of $b$, $i \in [ch(b)]$, the procedure recursively constructs the subset $C_i
= CV((T_{c_i(b)},c_i(b)),d)$, and then returns as output the vertex set
$\bigcup_{i=1}^{ch(b)} C_i \cup \{b\}$. (See Figure \ref{arbi} for an illustration.)

It is easy to see that the running time of this procedure is linear in $|T|$.

Let $(T,rt)$ be an $n$-vertex rooted tree, and let $d \ge 1$ be a fixed
parameter. For convenience, we define $n_i = |T_{c_i(b)}|$, for each $i \in [ch(b)]$.
Next, we analyze the properties of the set $C =
CV((T,rt),d)$ of \emph{cut vertices}.




Observe that for $n < 2d$, $C = \emptyset$,
and for $n \ge 2d$, $C$ is non-empty. \\
Next, we provide an upper bound on $|C|$ in the case $n \ge 2d$.
\begin {lemma} \label{boundingcv}
For $n \ge 2d$, $|C| \le (n/d) -1$.
\end{lemma}
\proof The proof is by induction on $n=|T|$. 
\\\emph{Basis: $2d \le n < 3d$.}
Fix an index $i \in [ch(b)]$. Since $b$ is balanced, we have
$$n_i ~\le~ n_1 ~\le~ n-d ~<~ 2d,$$ implying
that $C_i = \emptyset$. It follows that $C ~=~ \bigcup_{i=1}^{ch(b)}
C_i \cup \{b\} ~=~ \{b\},$ and so $|C| = 1 \le (n/d) -1.$
\\\emph{Induction Step:}
We assume the correctness of the statement for all smaller values of
$n$, $n \ge 3d$, and prove it for $n$. Let $I$ be the set of all indices $i$ in
$[ch(b)]$ for which $n_i \ge 2d$.
Observe that for each $i \in [ch(b)] \setminus I$, $C_i =
\emptyset$, and by the induction hypothesis, for each $i \in I$, $|C_i|
\le (n_i/d) -1$. By construction, the vertex sets
$C_1,C_2,\ldots,C_{ch(b)}$ and $\{b\}$ are pairwise disjoint, and $C
~=~ \bigcup_{i=1}^{ch(b)} C_i \cup \{b\}.$ Hence
\begin{equation} \label{sizeL}
|C|  ~=~ \sum_{i=1}^{ch(b)} |C_i| + 1 ~=~ \sum_{i \in I}
|C_i| + 1 ~\le~ \sum_{i\in I} ((n_i/d)-1) + 1.
\end{equation}
The analysis splits into three cases depending on the value of $|I|$.
\\\emph{Case 1: $|I|=0$.}
Equation (\ref{sizeL}) yields $|C| \le 1 \le (n/d) -1$.
\\\emph{Case 2: $|I|=1$.}
By construction, $n_1 \ge n_i$, for each $i \in [ch(b)]$, implying
that $I = \{1\}$. Since $b$ is balanced, $n_1 \le n-d$, and so
(\ref{sizeL}) yields
$$|C| ~\le~ (n_1/d) - 1 +1 ~\le~ (n-d)/d ~=~ (n/d) -1.$$
\\\emph{Case 3: $|I| \ge 2$.}
Clearly, $\sum_{i\in I} n_i  \le n-1$, and so (\ref{sizeL})
yields
$$|C| ~\le~ \sum_{i\in I} ((n_i/d)-1) + 1 ~=~ \sum_{i\in I} (n_i/d) - |I| + 1 ~\le~
(n-1)/d - 2 + 1 ~\le~ (n/d) -1.$$ \qed

Let $b=b(rt)$, and let $\overline {T_b}$ 
be the subtree of $T$ obtained by removing the subtree $T_b$ from $T$.
We use the following claim to prove Lemma \ref{sizetr}.
\begin {claim} \label{boundingtil}
$|\overline {T_b}| < d$.
\end {claim}
\begin{proof}
If $b = rt$, then $\overline {T_b}$ is empty and the assertion of the claim is immediate.
Otherwise, consider the parent $\pi(b)$ of $b$ in $T$. Since $b$ is
the first (i.e., closest to $rt$) balanced vertex along $P(rt)$, $\pi(b)$ is non-balanced,
and so $|T_b| = |T_{c_1(\pi(b))}| > n-d$. Hence $|\overline {T_b}| = n -
|T_b| <  d$, and we are done. \qed
\end{proof}

For a subset $U$ of $V(T)$, we denote by $T \setminus U$ the forest obtained from $T$ by removing all
vertices in $U$ along with the edges that are incident to them.
\begin{lemma} \label{sizetr}
The size of any subtree in the forest $T \setminus C$ is smaller than
$2d$.
\end {lemma}
\begin{proof}
The proof is by induction on $n = |T|$. The basis $n < 2d$ is
trivial.
\\\emph{Induction Step:} We assume the correctness of the statement for all smaller values of $n$, $n \ge 2d$,
and prove it for $n$. First, note that $b = b(rt) \in C$. 
Also, observe that for $n \ge 2d$, 
\begin{equation}
\label{decom} T \setminus C ~=~ \bigcup_{i=1}^{ch(b)} (T_{c_i(b)}
\setminus C_i) \cup \{\overline {T_b}\}. 
\end{equation}
Consider a subtree $T'$ in the forest $T \setminus C$. 
By (\ref{decom}), either $T' = \overline {T_b}$, or it belongs to the forest
$T_{c_i(b)} \setminus C_i$, for some index $i \in [ch(b)]$. In the
former case, the size bound follows from Claim \ref{boundingtil},
whereas in the latter case it follows from the induction hypothesis.
\qed
\end{proof}

Any subset $U$ of $V(T)$ induces a forest $\mathcal Q(T,U)$ over $U$ in the natural
way:
a vertex $v \in U$ is defined to be a child of its closest ancestor in $T$ that belongs to $U$.
Define $\mathcal Q = \mathcal Q(T,C)$. Observe that for $n < 2d$, $C = \emptyset$, and so $\mathcal Q = \emptyset$.
Also, for $n \ge 2d$, $C$ is non-empty and $b = b(rt) \ne NULL$.
\begin{lemma} \label{whatto}
For $n \ge 2d$, $\mathcal Q$ is a spanning tree of $C$ rooted at $b=b(rt)$, such
that for each vertex $v$ in $C$, the number of children
 of $v$ in $\mathcal Q$, denoted $ch_{\mathcal Q}(v)$, is no greater than the
corresponding number $ch(v)$ in $T$.
\end{lemma}
{\bf Remark:} This lemma implies that $\Delta(\mathcal Q) \le \Delta(T)$.
\begin{proof}
The proof is by induction on $n = |T|$.
\\\emph{Basis: $2d \le n < 3d.$}
In this case $C = \{b\}$, and so $\mathcal Q$ consists of a single
root vertex $b$.
\\\emph{Induction Step:} We assume the correctness of the statement for all smaller values of $n$, $n \ge 3d$,
and prove it for $n$. Let $I$ be the set of all indices $i$ in
$[ch(b)]$ for which $n_i \ge 2d$, and write $I =
\{{i_1},{i_2},\ldots,{i_{|I|}}\}$. Observe that for each index $i \in
[ch(b)] \setminus I$, $C_i = \emptyset$, and so $\mathcal Q(T_{c_i(b)},C_i)$ is an empty tree.
By the induction hypothesis, for each $i \in I$, $\mathcal Q_i =
\mathcal Q(T_{c_i(b)},C_i)$ is a spanning tree of $C_i$ rooted at
$b_i = b(c_i(b)) \ne NULL$ in which the number of children of each vertex is
no greater than the corresponding number in $T_{c_i(b)}$.
By definition, the only children of $b$ in $\mathcal Q$ are the
roots $b_{i_1},b_{i_2},\ldots,b_{i_{|I|}}$ of the non-empty trees
$\mathcal Q_{i_1},\mathcal Q_{i_2},\ldots,\mathcal Q_{i_{|I|}}$,
respectively,
and so $ch_{\mathcal Q}(b) = |I| \le ch(b)$. In addition, $b$ has no
parent in $\mathcal Q$, and so it is the root of $\mathcal Q$. \qed
\end{proof}

For a tree $\tau$, the root $rt(\tau)$ of $\tau$ and its left-most
vertex $l(\tau)$ are called the \emph{sentinels} of $\tau$.
The next lemma shows that each subtree in the forest $T \setminus C$
is incident to at most two cut vertices.
The proof of this lemma follows similar lines as those in the
proof of Lemma \ref{sizetr}, and is thus omitted. 
\begin{lemma} \label{connection}
For any subtree $T'$ in the forest $T \setminus C$, no other vertex in $T'$ other
than its two sentinels $rt(T')$ and $l(T')$ is incident to a vertex from $C$. Moreover,
both $rt(T')$ and $l(T')$ are incident to at most one vertex from $C$; specifically,
$rt(T')$ is incident to its parent in $T$, unless $rt(T')$ is the
root of $T$, and $l(T')$ is incident to its left-most child in $T$, unless $l(T')$ is a leaf in $T$.
\end {lemma}

\begin{figure*}[htp]
\begin{center}
\begin{minipage}{\textwidth} 
\begin{center}
\setlength{\epsfxsize}{4.5in} \epsfbox{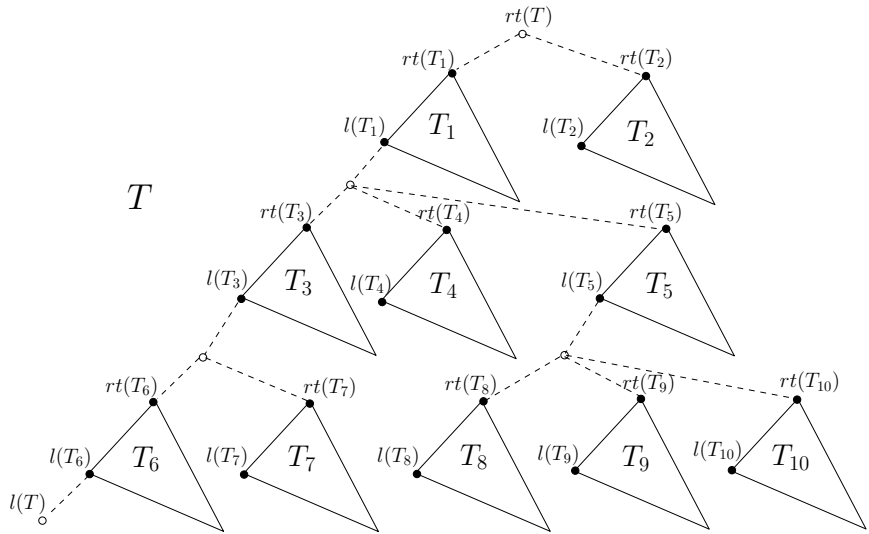}
\end{center}
\end{minipage}
\caption[]{ \label{dec} \sf \small A ``path-like'' decomposition of
the tree $T$ into subtrees $T_{1},T_2,\ldots,T_{10}$. 
The 5 cut vertices of $\tilde C$ (i.e., the 3
vertices of $C$ and the 2 sentinels $rt(T)$ and $l(T)$ of $T$) are depicted in the figure by empty dots, 
whereas the
20 sentinels of the subtrees $T_1,T_2,\ldots,T_{10}$
are depicted by filled dots. Similarly to the 1-dimensional case,
each subtree $T_i$ is incident to at most two cut vertices.
Edges in $T$ that connect sentinels of subtrees with cut vertices
are depicted by dashed lines.} 
\end{center}
\end{figure*}

%
Similarly to the 1-dimensional case, we add the two sentinels
$rt(T)$ and $l(T)$ of the original tree $T$ to the set $C$ of
cut vertices. From now on we refer to the appended set $\tilde {C} =
C \cup \{rt(T),l(T)\}$ as the set of \emph{cut vertices}.
Intuitively, Lemma \ref{connection} shows 
that the 
Procedure $CV$ 
``slices" the tree in a ``path-like"
fashion, i.e., in a way that is analogous to the decomposition of $\vartheta_n$ into intervals
described in Section \ref{sec221}. 
(See Figure \ref{dec} for an illustration.) 

Lemmas \ref{boundingcv}, \ref{sizetr}, \ref{whatto} and \ref{connection} imply
the following corollary, which summarizes the properties of the set $\tilde C$ of cut vertices.


\begin{corollary} \label{cen}
\begin{enumerate}
\item For $n \ge 2d$, $|\tilde{C}| \le (n/d) +1$.
\item The size of any subtree in the forest $T \setminus \tilde{C}$ is smaller than
$2d$.
\item $\tilde \mathcal Q = \mathcal Q(T,\tilde{C})$ is a
spanning tree of $\tilde {C}$ rooted at $rt(T)$, with $\Delta(\tilde \mathcal Q) \le \Delta(T)$.
\item For any
subtree $T'$ in the forest $T \setminus \tilde{C}$, only the two sentinels $rt(T')$ and $l(T')$ of $T'$ are incident to a vertex from
$\tilde{C}$. Moreover, both $rt(T')$ and $l(T')$ are incident to at most one vertex from $\tilde C$;
specifically,
$rt(T')$ is incident to its parent in $T$, and $l(T')$ is incident to its left-most child in $T$,
unless $l(T')$ is a leaf in $T$.
\end{enumerate}
\end{corollary}
{\bf Remark:} The running time of the Procedure $CV$ is $O(n)$.
Hence the set $\tilde C$ of cut vertices can be computed in linear time.

\subsubsection{1-Spanners with Low Diameter} \label{lowtrees} Consider
an $n$-vertex (weighted) tree $T$, and let $M_T$ be the tree metric
induced by $T$. In this section we devise a construction $\mathcal
H_k(n)$ of 1-spanners for $M_T$ with comparable monotone diameter
$\bar \Lambda(n) = \bar \Lambda(\mathcal H_k(n))$ in the range
$\Omega(\alpha(n)) = \bar \Lambda(n) = O(\log n)$. Both in this
construction and in the one of Section \ref{sec233}, all edges of
the original tree $T$ are added to the spanner.

Let $k$ be a fixed parameter such that $4 \le k \le n/2-1$, and set
$d=n/k$. (Observe that $n \ge 2k+2$ and $d > 2$.) To select the set $\tilde C$ of cut vertices, we invoke the 
procedure $CV$ on the input $(T,rt)$ and $d$. Set $C = CV((T,rt),d)$ and $\tilde{C} = C \cup
\{rt(T),l(T)\}$. Since $k \ge 4$, it holds that $2d = 2n/k
< n$. Denote the subtrees in the forest $T \setminus \tilde{C}$ by
$T_1,T_2,\ldots,T_p$. By Corollary \ref{cen},
$|\tilde{C}| \le (n/d)+1 = k+1$, and each subtree $T_i$ in $T \setminus
\tilde{C}$ has size less than $2d=2n/k$. Observe that
$\sum_{i=1}^p |T_i| ~=~ n - |\tilde{C}| ~\ge~ n-k-1,$ implying
that the number $p$ of subtrees in $T \setminus \tilde{C}$ satisfies
\begin{equation} \label{pvs}
p ~\ge~ \frac{n-k-1}{2n/k} ~\ge~ k/4.
\end{equation}
(The last inequality holds for $k \le n/2-1$.)

To connect the set $\tilde{C}$ of cut vertices,
the algorithm first constructs the tree $\tilde \mathcal Q = \mathcal
Q(T,\tilde{C})$. 
Observe that $\tilde \mathcal Q$ inherits the tree structure of $T$, that is, for any two points $u$ and $v$ in
$\tilde{C}$, $u$ is an ancestor of $v$ in $\tilde \mathcal Q$ if and only if it is
its ancestor in $T$. Consequently, any 1-spanner path for the tree metric $M_{\tilde \mathcal Q}$ induced by $\tilde \mathcal Q$
between two arbitrary comparable\footnote{This
may not hold true for two points that are not comparable, as their
least common ancestor may not belong to $\tilde \mathcal Q$.} points is also a 1-spanner
path for the original tree metric $M_T$.
The algorithm proceeds by building a 1-spanner for $\tilde \mathcal
Q$ via one of the aforementioned generalized constructions from
\cite{Chaz87,AS87,Thor97,Sol11} (henceforth, \emph{tree-spanner}).
In other words, $O(k)$ edges between cut vertices are added to the
spanner $\mathcal H_k(n)$ to guarantee that the monotone distance in
the spanner between any two comparable cut vertices
will be $O(\alpha(k))$.
Then the algorithm adds to the spanner $\mathcal H_k(n)$ edges that connect each of the two sentinels to all other
cut vertices. (In fact, the leaf $l(T)$ needs not be connected to all
cut vertices, but rather only to those which are its ancestors in
$T$.) Finally, the algorithm calls itself recursively for each of
the subtrees $T_1,T_2,\ldots,T_p$ of $T$. At the bottom level of the
recursion, i.e., when $n < 2k+2$, the algorithm uses the
tree-spanner to connect all points, and, in addition, it adds to the spanner edges that connect
each of the two sentinels $rt(T)$ and $l(T)$ to all the other $n-1$ points.


We denote by $E(n)$ the number of edges in $\mathcal H_k(n)$,
excluding edges of $T$. Clearly, $E(n)$
satisfies the recurrence $E(n) \le O(k) + \sum_{i=1}^{p}
E(|T_i|)$, with the base condition $E(q) =O(q)$, for all $q <
2k+2$. Recall that for each
$i \in
[p]$, $|T_i| \le 2d=2n/k < n$,
and by Equation (\ref{pvs}), we have $p \ge k/4$.
Also, since $\widetilde {C}$ is non-empty,
it holds that $\sum_{i=1}^p |T_i| = n-|\tilde{C}| \le n-1$. 
Next, we prove by induction on $n$ that $E(n) \le 4c(n-1)$, for a sufficiently large constant $c$. 
The basis $n < 2k+2$ is immediate. For $n \ge 2k+2$,
the induction hypothesis implies that
$$E(n) ~\le~ c \cdot k + 4c\cdot \sum_{i=1}^p (|T_i| - 1) ~=~ c \cdot k
- 4c \cdot p + 4c \cdot \sum_{i=1}^p |T_i| ~\le~ c (k -4p) +
4c (n-1) ~\le~ 4c(n-1).$$ (The last inequality holds as $p \ge k/4$.)

Denote by $\Delta(n)$ the maximum degree of a
vertex in $\mathcal H_k(n)$, excluding edges of $T$. Since
$|\tilde{C}| \le k+1$, $\Delta(n)$ satisfies the recurrence
$\Delta(n) \le \max\{k,\Delta(2n/k)\}$, with the base condition $\Delta(q)
\le 2k$, for all $q < 2k+2$. It follows that $\Delta(n) \le 2k$.
Including edges of the tree $T$, the number of edges increases by at most $n-1$ units
and the maximum degree increases by at most $\Delta(T)$ units.


Next, we show that the lightness $\Psi(\mathcal H_k(n))$ of the
spanner $\mathcal H_k(n)$ satisfies $\Psi(\mathcal H_k(n)) = O(k
\cdot \log_k n)$. To this end, we extend the notion of $\emph{load}$
defined in \cite{AWY05}\footnote{Agarwal et al.\ \cite{AWY05} used
a slightly different notion which they called \emph{covering}. The
notion of load as defined above was introduced in \cite{DES09tech}, but
the two notions are very close.} for 1-dimensional spaces to general
tree metrics. Consider an edge $e' = (v,w)$ connecting two arbitrary
points in $M_T$, and an edge $e \in E(T)$. The edge $e'$ is said
to \emph{load} $e$ if the unique path in $T$ between the endpoints
$v$ and $w$ of $e'$ traverses $e$. For a spanning subgraph $H$ of
$M_T$, the number of edges $e' \in E(H)$ that load an edge $e \in E(T)$
is called the \emph{load} of $e$ by $H$, and is denoted by $\chi(e) =
\chi_H(e)$. The \emph{load} of $H$ (with respect to $T$), $\chi(H) = \chi_T(H)$, is the maximum load of 
an edge of $T$ by $H$. By double-counting,
\begin{eqnarray} \label{loadeq} \nonumber w(H) &=& \sum_{e' \in E(H)} w(e')
~=~ \sum_{e' \in E(H)} \sum_{\{e \in E(T)~:~ e \mbox{\footnotesize { loaded by }}e'\}} w(e) 
~=~ \sum_{e \in E(T)} \sum_{\{e' \in E(H)~:~ e' \mbox{\footnotesize { loads }} e\}} w(e)
\\ &=& \sum_{e \in E(T)} \chi_H(e) \cdot w(e) ~\le~ \chi(H) \cdot
\sum_{e \in E(T)} w(e) ~=~ \chi(H) \cdot w(T), \end{eqnarray}
implying that $\Psi(H) = w(H)/w(T) \le \chi(H)$. Thus it suffices
to provide an upper bound of $O(k \cdot \log_k n)$ on the load $\chi(\mathcal H_k(n))$
of
$\mathcal H_k(n)$. Denote by $\chi(n)$ the load of $\mathcal H_k(n)$, excluding edges of $T$.
After the first level of recursion, 
$\mathcal H_k(n)$ contains only $O(k)$ edges that connect
cut vertices.
These edges contribute $O(k)$ units of load to each edge of $T$. In particular, after the first level of recursion,
each subtree in the forest $T \setminus \tilde C$ is loaded by at most $O(k)$ edges.  
Hence $\chi(n)$ satisfies the
recurrence $\chi(n) \le O(k) + \chi(2n/k)$, with the base condition
$\chi(q) = O(q)$, for all $q < 2k+2$, yielding $\chi(n) = O(k \cdot \log_k n)$.
Including edges of the tree $T$, the load increases by one unit, and we are done.

Next, we show that $\bar \Lambda(n) = \bar \Lambda(\mathcal H_k(n))
= O(\log_k n + \alpha(k))$. The \emph{leaf radius} $\hat R(n)$ of
$\mathcal H_k(n)$ is defined as the maximum monotone distance
between the left-most vertex $l(T)$ in $T$ and one of its ancestors
in $T$. By Corollary \ref{cen}, similarly to the 1-dimensional case,
$\hat R(n)$ satisfies the recurrence  $\hat R(n) \le 2 + \hat
R(2n/k)$, with the base condition $\hat R(q) = 1$, for all $q < 2k+2$.
Hence, $\hat R(n) = O(\log_k n)$. Similarly, we define the
\emph{root radius} $\check R(n)$ as the maximum monotone distance
between the root $rt(T)$ of $T$ and some other point in $T$. By the
same argument
 we get $\check R(n) = O(\log_k n)$.
Applying again Corollary \ref{cen} and reasoning similar to the 1-dimensional case, we get that $\bar \Lambda(n) \le
\max\{\bar \Lambda(2n/k),O(\alpha(k))+\check R(2n/k) + \hat R(2n/k)\}$,
with the base condition $\bar \Lambda(q) = O(\alpha(q))$, for all $q < 2k+2$. It follows that $\bar \Lambda(n) = O(\log_k n + \alpha(k))$.

Finally, we argue that the worst-case running time of the algorithm, denoted $t(n)$,  is $O(n \cdot \log_k n)$.
The algorithm starts by invoking the decomposition procedure
for selecting the set $\tilde C$ of cut vertices. As was mentioned above, this step requires $O(n)$ time.
Next, the algorithm builds the tree $\tilde Q$, which
can be carried out in time $O(|\tilde Q|) = O(k)$. The algorithm proceeds by building the tree-spanner
for $\tilde Q$. The tree-spanner of \cite{Chaz87,AS87,Thor97,Sol11} can be built within linear time.
Hence, building the tree-spanner for $\tilde Q$ requires $O(k)$ time. Next, the algorithm adds to the spanner edges that connect
each of the two sentinels to all other cut vertices, which can be carried out within time $O(k)$ as well. Finally,
the algorithm calls itself recursively for each of the subtrees $T_1,T_2,\ldots,T_p$ of $T$,
which requires at most $\sum_{i=1}^p t(|T_i|)$ time.
At the bottom level of the recursion, i.e., when $n < 2k+2$, the algorithm uses the tree-spanner
to connect all points, and in addition, it adds to the spanner edges that connect each of the two
sentinels of the tree to all the other $n-1$ points. 
Hence, the running time of the algorithm at the bottom level of the recursion is $O(n)$.
It follows that  $t(n)$ satisfies 
the recurrence $t(n) \le O(n) + \sum_{i=1}^p t(|T_i|)$, with the base condition
$t(q) = O(q)$, for all $q < 2k+2$. Recall that $k \ge 4$, and for each $i \in [p]$, $|T_i| \le 2n/k < n$.
We conclude that $t(n) = O(n \cdot \log_k n)$.
\begin{theorem} \label{thmlow} For any
tree metric $M_T$ and a parameter $k$, there exists a
1-spanner with maximum degree at most $\Delta(T) + 2k$, diameter
$O(\log_k n + \alpha(k))$, lightness $O(k \cdot \log_k n)$, and $O(n)$ edges.
The running time of this construction is $O(n \cdot \log_k n)$.
\end{theorem}
We remark that the maximum degree $\Delta(\mathcal H)$ of the spanner $\mathcal H = \mathcal H_k(n)$ cannot be in general
smaller
than the maximum degree $\Delta(T)$ of the original tree. Indeed, consider a unit weight star $T$ with edge set $\{(rt,v_1),(rt,v_2),\ldots,(rt,v_{n-1})\}$.
Obviously, any spanner $\mathcal H$ for $M_T$ with $\Delta(\mathcal H) < n-1$ distorts
the distance between the root $rt$ and some other vertex.
\subsubsection{1-Spanners with High Diameter} \label{sec233}
In this section we devise a construction $\mathcal
H'_k(n)$ of 1-spanners for $M_T$ with comparable monotone diameter
$\bar \Lambda'(n) = \bar \Lambda(\mathcal H'_k(n))$ in the range
$\bar \Lambda'(n) = \Omega(\log n)$. 



The algorithm starts with constructing the tree $\tilde \mathcal Q =
\mathcal Q(T,\tilde{C})$ that spans the set $\tilde{C}$ of
cut vertices. All edges of $\tilde \mathcal Q$ are inserted into $\mathcal H'_k(n)$.
(This step is analogous to taking the edges $(r_0,r_1),(r_1,r_2),\ldots,(r_{k-1},r_k)$ 
in the 1-dimensional construction of Section \ref{sec223}.)
Observe that the depth of $\tilde \mathcal Q$ is at
most $k$, implying that any two comparable cut vertices are
connected via a 1-spanner path in
$\tilde \mathcal Q$ that consists
of at most $k$ edges; since $\tilde \mathcal Q$ inherits the tree
structure of $T$, this path is also a 1-spanner path for the original tree metric $M_T$.
Then the algorithm calls itself recursively for each of the subtrees
$T_1,T_2,\ldots,T_p$ of $T$. At the bottom level of the recursion, i.e.,
when $n < 2k+2$, the algorithm adds no additional edges to the spanner.

Similarly to Section \ref{lowtrees} it follows that the number of edges in
$\mathcal H'_k(n)$ is $O(n)$.
Next, we analyze the maximum degree of this construction.
Denote by $\Delta'(n)$ the maximum degree of a vertex in $\mathcal
H'_k(n)$, excluding edges of $T$,
and let $\Delta_0 = \Delta(T)$ denote the maximum degree of the original tree $T$.
By the third assertion of Corollary
\ref{cen}, $\Delta(\tilde \mathcal Q) \le \Delta_0$, and so
$\Delta'(n)$ satisfies the recurrence $\Delta'(n) \le
\max\{\Delta_0,\Delta'(2n/k)\}$, with the base condition
$\Delta'(q) = 0$, for all $q < 2k+2$, yielding $\Delta'(n) \le
\Delta_0$.  It follows that the maximum degree $\Delta(\mathcal
H'_k(n))$ of $\mathcal H'_k(n)$ is at most $2 \cdot \Delta_0 = 2 \cdot \Delta(T)$.


Next, we show that the load $\chi(\mathcal H'_k(n))$
of $\mathcal H'_k(n)$ is $O(\log_k n)$, which, by (\ref{loadeq}), implies
that $\Psi(\mathcal H'_k(n)) = O(\log_k n)$.
Denote by $\chi'(n)$ the load of $\mathcal H'_k(n)$, excluding edges of $T$.
After the first level of recursion, $\mathcal H'_k(n)$ contains just the edges of the tree $\tilde \mathcal Q$.
We argue that 
each edge $e=(u,v)$ of $T$ 
is loaded by at most one edge of $\tilde \mathcal Q$.
Indeed, if both $u$ and $v$ are cut vertices, then $e$ is also an edge of $\tilde \mathcal Q$, and so it is loaded by itself.
Otherwise, either $u$ or $v$ (or both of them) belongs to some subtree $T_i$ in the forest $T \setminus \tilde C$.
In this case, the fourth assertion of Corollary \ref{cen} implies that
$e$ is loaded by at most one edge in $\tilde \mathcal Q$,
namely, the edge connecting the parent of $rt(T_i)$ in $T$ and the
left-most child of $l(T_i)$ in $T$, if exists. 
In particular, after the first level of recursion, each subtree in the forest $T \setminus
\tilde{C}$ is loaded by at most one edge.
Hence $\chi'(n)$ satisfies the recurrence $\chi'(n) \le 1 +
\chi'(2n/k)$, with the base condition $\chi'(q) = 0$, for all $q <
2k+2$. It follows that $\chi'(n) = O(\log_k n)$.
Including edges of $T$, the load increases by one unit, and we are done.


By Corollary \ref{cen}, similarly to the 1-dimensional case, the leaf
radius $\hat R'(n)$ of $\mathcal H'_k(n)$ satisfies the recurrence
$\hat R'(n) \le k + \hat R'(2n/k)$, with the base condition $\hat
R'(q) \le q-1$, for all $q < 2k+2$, yielding $\hat R'(n) = O(k \cdot
\log_k n)$. Similarly, we get that $\check R'(n) = O(k \cdot \log_k
n)$. Applying Corollary \ref{cen} and reasoning similar to the
1-dimensional case, we get that the comparable monotone diameter
$\bar \Lambda'(n) = \bar \Lambda(\mathcal H'_k(n))$ of $\mathcal
H'_k(n)$ satisfies the following recurrence $\bar \Lambda'(n) \le
\max\{\bar \Lambda'(2n/k), k+ \check R'(2n/k) + \hat R'(2n/k)\}$,
with the base condition $\bar \Lambda'(q) \le q-1$, for all $q < 2k+2$.
It follows that $\bar \Lambda'(n) = O(k \cdot \log_k n)$.

We remark that $\mathcal H'_k(n)$ is a planar graph.  

Finally, by employing an argument very similar to the one 
used in Section \ref{lowtrees},
we get that the worst-case running time of the algorithm 
is  $O(n \cdot \log_k  n)$.

\begin{theorem}  \label{thmhigh} For any
tree metric $M_T$ and a parameter $k$, there exists a
1-spanner with maximum degree at most $2 \cdot \Delta(T)$, diameter $O(k
\cdot \log_k n)$, lightness $O(\log_k n)$, and $O(n)$ edges. 
Moreover, this 1-spanner is a planar graph.
The running time of this construction is $O(n \cdot \log_k n)$.
\end{theorem}

\section{Euclidean Spanners} \label{imp}
In this section we demonstrate that our 1-spanners for tree metrics can be used
for constructing Euclidean spanners and spanners for doubling metrics.

We start with employing the Dumbbell Theorem of \cite{ADMSS95} in conjunction with our 1-spanners for tree metrics to construct
 Euclidean spanners. 
\begin{theorem} (``Dumbbell Theorem'', Theorem 2 in \cite{ADMSS95}) Given a set $S$ of $n$ points in
$\mathbb R^d$ and a parameter $\eps > 0$, a forest
$\mathcal D$ consisting of $O(1)$ rooted binary trees of size $O(n)$ each can be built in time $O(n \cdot \log n)$, having the
following properties:
\begin{enumerate}
\item For each tree in $\mathcal D$, there is a 1-1 correspondence
between the leaves of this tree and the points of $S$. 
\item Each
internal vertex in the tree has a unique representative point, which
can be selected arbitrarily from the points in any of its descendant
leaves. 
\item Given any two points $u,v \in S$, there is a tree in
$\mathcal D$, so that the path formed by walking from representative
to representative along the unique path in that tree between these
vertices, is a $(1+\eps)$-spanner path for $u$ and $v$.
\end{enumerate}
\end{theorem}

For each dumbbell tree in $\mathcal D$, we use the following
representative assignment from \cite{ADMSS95}. Leaf labels are
propagated up the tree. An internal vertex chooses to itself one of
the propagated labels and propagates the other one up the tree. Each
label is used at most twice, once at a leaf and once at
an internal vertex. 
Any label assignment induces a weight function over the edges of the dumbbell tree in the obvious way.
(The weight of an
edge is set to be the Euclidean distance between the representatives
corresponding to the two endpoints of that edge.) Arya et
al.\ \cite{ADMSS95} proved that the lightness of dumbbell trees is
always $O(\log n)$, regardless of which
representative assignment is chosen for the internal vertices.

Next, we describe our construction of Euclidean spanners with
diameter in the range $\Omega(\alpha(n)) = \Lambda = O(\log n)$.

We remark that each dumbbell tree has size $O(n)$. 
For each
(weighted) dumbbell tree $DT_i \in \mathcal D$, denote by $M_{i}$
the $O(n)$-point tree metric induced by $DT_i$. To obtain our
 construction of $(1+\eps)$-spanners with low diameter, we set $k =
n^{1/\Lambda}$, and build the 1-spanner construction $\mathcal H^i =
\mathcal H^i_k(O(n))$ that is guaranteed by Theorem \ref{thmlow} for each of the tree
metrics $M_{i}$. Then we translate each $\mathcal H^i$ to be a
spanning subgraph $\breve \mathcal H^i$ of $S$ in the obvious way.
(Each edge in $\mathcal H^i$ is replaced with an edge that connects
the representatives 
corresponding to the endpoints of that edge.)
Finally, let $\mathcal E_k(n)$ be the spanner obtained from the union of all the graphs $\breve \mathcal H^i$.

Theorem \ref{thmlow}
implies that each graph $\breve \mathcal H_i$ contains only $O(n)$ edges.
By the Dumbbell Theorem, $\mathcal E_k(n)$ is the union of  a constant number of such graphs.
Thus the total number of edges in $\mathcal E_k(n)$ is $O(n)$.

We proceed by showing that $\mathcal E_k(n)$ is a $(1+\eps)$-spanner for $S$ with
diameter
$\Lambda = \Lambda(\mathcal E_k(n))$ at most $O(\log_k n +
\alpha(k))$. By the Dumbbell Theorem, for any pair $u,v$  of points in $S$, there is a dumbbell tree $DT_i$, so that the unique
path $P_{u,v}$ connecting $u$ and $v$ in $DT_i$ is a
$(1+\eps)$-spanner path for them. Theorem \ref{thmlow} implies that
there is a 1-spanner path $P$ in $\mathcal H^i$ between $u$ and $v$
that consists of at most $O(\log_k n + \alpha(k))$ edges. By the
triangle inequality, the weight of the corresponding translated path
$\breve P$ in $\breve \mathcal H^i$ is no greater than the weight of
$P_{u,v}$. Hence, $\breve P$ is a $(1+\eps)$-spanner path for $u$ and
$v$ that consists of at most $O(\log_k n + \alpha(k))$ edges.

Next, we show that the maximum degree $\Delta(\mathcal E_k(n))$ of $\mathcal E_k(n))$ is $O(k)$. 
Since
each dumbbell tree $DT_i$ is binary, Theorem \ref{thmlow} implies
that $\Delta(\mathcal H^i) = O(k)$. Recall that each label is used
at most twice in $DT_i$, and so $\Delta(\breve \mathcal H^i) \le 2
\cdot \Delta(\mathcal H^i) = O(k)$.
The union of $O(1)$ such graphs will also have maximum degree $O(k)$.

We argue that the lightness $\Psi(\mathcal E_k(n))$ of $\mathcal E_k(n)$ is $O(k \cdot \log_k n
\cdot \log n)$. 
Consider an arbitrary dumbbell tree $DT_i$. Recall that the lightness of all dumbbell trees
is $O(\log n)$, and so $w(DT_i) = O(\log n) \cdot w(MST(S))$. By Theorem \ref{thmlow}, the
weight $w(\mathcal H^i)$ of $\mathcal H^i$ is at most
$O(k \cdot
\log_k n) \cdot w(DT_i) = O(k \cdot \log_k n \cdot \log
n) \cdot w(MST(S))$. By the triangle inequality, the weight of
each edge in $\breve \mathcal H^i$ is no greater than the
corresponding weight in $\mathcal H^i$, implying that
the weight 
$w(\breve \mathcal H_i)$
of the graph $\breve \mathcal H_i$
satisfies
$w(\breve
\mathcal H^i) \le w(\mathcal H^i) = O(k \cdot \log_k n \cdot \log
n) \cdot w(MST(S)).$
The union of $O(1)$ such graphs will also have
weight $O(k \cdot \log_k n \cdot \log n) \cdot w(MST(S))$.

Finally, we bound the running time of this construction.
By the Dumbbell Theorem, the forest $\mathcal D$ of dumbbell trees can be built in $O(n \cdot \log n)$ time.
Theorem \ref{thmlow} implies that we can compute each of the graphs $\mathcal H^i$ in time $O(n \cdot \log_k n) = O(n \cdot \log n)$.
Moreover, as each graph $\mathcal H^i$ contains only $O(n)$ edges, translating it
into a graph $\breve \mathcal H^i$ as described above 
can be carried out in $O(n)$ time.
Since there is a constant number of such graphs, it follows that the overall time needed to compute our
construction $\mathcal E_k(n)$ of Euclidean spanners is $O(n \cdot \log n)$.

To obtain our construction of Euclidean spanners for the complementary range $\Lambda = \Omega(\log n)$, we
use our 1-spanners for tree metrics from Theorem \ref{thmhigh} instead of Theorem \ref{thmlow}.
\begin{corollary}
For any set $S$ of $n$ points in $\mathbb R^d$, any $\eps
> 0$ and a parameter $k$, there exists a
$(1+\eps)$-spanner with maximum degree $O(k)$, diameter $O(\log_k n + \alpha(k))$, lightness $O(k \cdot
\log_k n \cdot \log n)$, and $O(n)$ edges. There also exists
a $(1+\eps)$-spanner with maximum degree $O(1)$, diameter $O(k \cdot
\log_k n)$, and lightness $O(\log_k n \cdot \log n)$. 
Both these constructions can be implemented in time $O(n \cdot \log n)$.
\end{corollary}

In Appendix \ref{constlight} we show that the lightness of well-separated pair constructions 
for random point sets in the unit cube is (w.h.p.) $O(1)$. Also, the lightness
of well-separated pair constructions provides an asymptotic upper bound on the lightness of dumbbell trees. 
We derive 
the following result as a  corollary.
\begin{corollary}
For any set $S$ of $n$ points that are chosen independently and uniformly at random from the unit cube, any $\eps
> 0$ and a parameter $k$, there exists a
$(1+\eps)$-spanner with 
maximum degree $O(k)$, diameter $O(\log_k n + \alpha(k))$, lightness (w.h.p.) $O(k \cdot
\log_k n)$, and $O(n)$ edges. There also exists
a $(1+\eps)$-spanner with maximum degree $O(1)$, diameter $O(k \cdot
\log_k n)$, and lightness (w.h.p.) $O(\log_k n)$. 
Both these constructions can be implemented in time $O(n \cdot \log n)$.
\end{corollary}

Arya et al.\ \cite{ADMSS95} devised a well-separated pair construction of $(1+\eps)$-spanners with both diameter and lightness
at most $O(\log n)$. 
In addition, Lenhof et al.\ \cite{LSW94} showed that there exist
point sets for which any well-separated pair construction must admit lightness at least $\Omega(\log n)$.
While this existential bound holds true in the worst-case scenario, our probabilistic upper bound of $O(1)$ on the lightness of 
well separated pair constructions for random point sets implies that on average one can do much better.
\begin{corollary}
For any set $S$ of $n$ points that are chosen independently and uniformly at random from the unit cube, 
there exists a $(1+\eps)$-spanner with 
diameter $O(\log n)$, lightness (w.h.p.) $O(1)$, and $O(n)$ edges.
This construction can be implemented in $O(n \cdot \log n)$ time.
\end{corollary}


Chan et al.\  \cite{CGMZ05} showed that for any doubling metric $(X,\delta)$ there exists a $(1+\eps)$-spanner
with constant maximum degree. On the way to this result they proved the
following lemma, which we employ in conjunction with our 1-spanners for tree metrics
to construct spanners for doubling metrics.
\begin{lemma} [Lemma 3.1 in \cite{CGMZ05}] \label{Chanet}
For any doubling metric $(X,\delta)$, there exists a collection $\mathcal T$ 
of $m=O(1)$ spanning trees for $(X,\delta)$, $\mathcal T = \{\tau_1,\tau_2,\ldots,\tau_m\}$, that satisfies the following two properties:
\begin{enumerate}
\item For each index $i \in [m]$, the maximum degree $\Delta(\tau_i)$ of the tree $\tau_i$ is constant,
i.e., $\Delta(\tau_i) = O(1)$.
\item For each pair of points $x, y \in X$ there exists an index $i \in [m]$, such that 
$dist_{\tau_i}(x,y) = O(1) \cdot \delta(x,y)$.
\end{enumerate}
\end{lemma}


To obtain our spanners for doubling metrics we start with constructing the collection $\mathcal T = \{\tau_1,\tau_2,\ldots,\tau_m\}$
of spanning trees with properties listed in Lemma \ref{Chanet}.  
Next, we apply Theorem \ref{thmlow} with some parameter
$k$ to construct a $1$-spanner $\mathcal Z^i = \mathcal Z^i_k(n)$ for the tree metric induced by
the $i$th tree $\tau_i$ in $\mathcal T$, for each $i \in [m]$. Notice that, in general, edge weights in the graphs
$\mathcal Z^i$, $i \in [m]$, may be greater than the corresponding metric distances; 
for each $i \in [m]$,
let $\breve \mathcal Z^i$ be the graph obtained from $\mathcal Z^i$, 
by assigning weight $\delta(x,y)$ to each edge $(x,y) \in \mathcal Z^i$.
Our 
spanner $\mathcal Z$ is set to be the union of all the graphs $\breve \mathcal Z_i$, i.e., 
$\mathcal Z = \bigcup_{i=1}^m \breve \mathcal Z_i$.


By Theorem \ref{thmlow}, each of the graphs $\breve \mathcal Z^i$ contains only $O(n)$ edges.
Hence, the number of edges in $\mathcal Z$ is at most $m \cdot O(n) = O(n)$.

To argue that $\mathcal Z$ is an $O(1)$-spanner for $(X,\delta)$ consider a pair of points $x,y \in X$. By Lemma \ref{Chanet}, there exists
an index $i \in [m]$, such that $dist_{\tau_i}(x,y) = O(1) \cdot \delta(x,y)$. Since $\mathcal Z^i$ is a $1$-spanner
for the metric induced by $\tau_i$, it follows that $dist_{\mathcal Z^i}(x,y) = dist_{\tau_i}(x,y)$. 
Also, we have $dist_{\breve \mathcal Z^i}(x,y) \le dist_{\mathcal Z^i}(x,y)$.
Finally, since $\breve \mathcal Z^i \subseteq \mathcal Z$, we conclude
that $dist_{\mathcal Z}(x,y) \le dist_{\breve \mathcal Z^i}(x,y) \le dist_{\mathcal Z^i}(x,y) = dist_{\tau_i}(x,y) = O(1) \cdot \delta(x,y)$.
Observe also that $\Lambda(\mathcal Z^i) = O(\log_k n + \alpha(k))$, and so there is a path between $x$ and $y$ in $\mathcal Z^i$
that consists of at most $\Lambda(\mathcal Z^i)$ edges and has length at most $dist_{\mathcal Z^i}(x,y)$. Consequently,
$\Lambda(\mathcal Z) = O(\log_k n + \alpha(k))$. 

By Theorem \ref{thmlow}, the maximum degree of each graph $\breve \mathcal Z^i$ satisfies $\Delta(\breve \mathcal Z^i) \le \Delta(\tau_i) + 2k$.
By Lemma \ref{Chanet}, for each index $i \in [m]$, $\Delta(\tau_i) = O(1)$. Hence $\Delta(\breve \mathcal Z^i) = O(k)$. Since $m= O(1)$,
it follows that $\Delta(\mathcal Z) \le \sum_{i=1}^m \Delta(\breve \mathcal Z^i) \le m \cdot O(k) = O(k)$, and we are done.

\begin{corollary} \label{chinchan}
For any $n$-point doubling metric $(X,\delta)$ and a parameter $k$, there exists
an $O(1)$-spanner with maximum degree $O(k)$, diameter $O(\log_k n + \alpha(k))$, and $O(n)$ edges.
\end{corollary}


\section{Acknowledgments} 
We are grateful to Sunil Arya, David Mount and Michiel Smid for
helpful discussions.

\clearpage

\clearpage
\pagenumbering{roman}
\appendix
\centerline{\LARGE\bf Appendix}


\section{Well-Separated Pair Constructions for Random Point Sets} \label{constlight}
In this appendix we show that for any set 
$\mathcal S$ of points that are chosen independently and uniformly at random from 
the unit square, the lightness of well-separated pair constructions is (w.h.p.) $O(1)$.
Our argument also extends to higher constant dimensions.

The following lemma from \cite{NS07} provides a lower bound on the weight of $MST(\mathcal S)$. 
\begin{lemma} [Lemma 15.1.6 in \cite{NS07}]
For a set $\mathcal S$ of $n$ points that are chosen independently and uniformly at random from 
the unit square, there are constants $c > 0$ and $0 < \rho < 1$, such that
\\$Pr(w(MST(\mathcal S)) < c \cdot \sqrt{n}) \le \rho^n$. 
\end{lemma}

The following statement shows that the lightness of well-separated pair constructions 
for $\mathcal S$ is (w.h.p.) $O(1)$.
\begin{proposition} \label{maini}
For \emph{any} set $S$ of $n$ points in the unit square, the weight of well-separated
pair constructions is 
$O(\sqrt{n})$.
\end{proposition}
Before we prove Proposition \ref{maini}, we provide (Appendix \ref{backnot}) 
the relevant background and introduce some notation.
 The proof of Proposition \ref{maini} appears in Appendix \ref{arg}.  

{\bf Remark:} 
After communicating this result to Michiel Smid, he 
\cite{Smid09} pointed out the following alternative argument for obtaining this  
probabilistic bound of $O(1)$ on the lightness of well-separated pair constructions.
First, Chandra \cite{Chandra94} showed that for random point sets in the unit cube,
any edge set that satisfies the gap property has lightness (w.h.p.) $O(1)$.
Second, consider the edge set $E$ of the dumbbell trees of \cite{ADMSS95}. 
As shown in \cite{ADMSS95} this set can be partitioned into
$E = E' \cup E''$, such that $E'$ satisfies the \emph{gap property} and $w(E'') = O(w(E'))$.
Finally, use the observation that the lightness
of well-separated pair constructions is asymptotically equal to that of dumbbell trees.  
On the other hand, our proof   employs a simple, self-contained, combinatorial argument for analyzing
the lightness of well-separated pair constructions \emph{directly}. Hence we believe that our approach is 
advantageous, since, in particular, it does not take a detour through the heavy dumbbell trees machinery
of \cite{ADMSS95}.

\ignore{

\begin{table}
\begin{center}
\resizebox {\textwidth}{!}{
\begin{tabular}{|c|c|c|c|c|c|c|c|c|c|c|c|c|}
\hline
 & \cite{ADMSS95}  & \cite{ADMSS95} & \cite{DN94}  & \cite{ADMSS95} &  {\bf New} & {\bf New} & {\bf New} &{\bf New}
&{\bf New} &{\bf New} &{\bf New}  & {\bf New} \\
\hline
 &  & & & & {\bf I,II} & {\bf I} & {\bf I} & {\bf I} & {\bf II} & {\bf II}
& {\bf II} & \\
\hline  $k$ &  & & & & 1 & $\log^\delta n$ & $2^{\sqrt{\log n}}$ &
$n^{1/\alpha(n)}$ & $\log^\delta n$ & $2^{\sqrt{\log n}}$
& $n^\zeta$ & \\
\hline \hline $\Delta$ & 1    & $n$ & 1 & $n$ & $1$ & $\log^\delta n$ &
$2^{\sqrt{\log n}}$ & $n^{1/\alpha(n)}$  & 1 & 1 & 1 & $n$ \\
\hline $\Lambda$ & $\log n$   &  $\alpha(n)$ & $n$ & $\log n$ & $\log n$ & ${{\log n} \over
{\log\log n}}$ & $\sqrt{\log n}$  & $\alpha(n)$  & $\frac{\log^{1+
\delta} n}{\log \log n}$ & $2^{O(\sqrt{\log
n})}$ & $n^\zeta$ & $\log n$ \\
\hline $\Psi$    &  $\log^2 n$   & $n$ & $1$ & $\log n$ & $\log n$ & ${{\log^{1 + \delta}
n} \over {\log\log n}}$    & $2^{O(\sqrt{\log n})}$
& $n^{O(1/\alpha(n))}$ &${{ \log n} \over {\log\log n}} $ & $\sqrt{\log n}$ & $1$ & $1$ \\
\hline
\end{tabular}
}
\end{center}
\caption[]{ \label{tab2} \footnotesize A concise comparison of previous and new
results for random point sets in the unit cube. (The previous results
hold for arbitrary point sets.)
Each column corresponds to a set of parameters that can be
achieved simultaneously. For each column the first row indicates
whether the result is new or not. 
For new results, the second row indicates whether it
is obtained by the first (I) or the second (II) tradeoff. The right-most column
is a new result that is not obtained by any of these two tradeoffs. 
(The first tradeoff is degree $O(k)$, diameter $O(\log_k n + \alpha(k))$, and
lightness $O(k \cdot \log_k n)$. The second tradeoff is
degree $O(1)$, diameter $O(k \cdot \log_k n)$ and lightness
$O(\log_k n)$.) The third row indicates the value of
$k$ that is substituted in the corresponding tradeoff. The next
three rows indicate the resulting degree ($\Delta$), diameter
$(\Lambda)$ and lightness $(\Psi)$. The bounds on lightness in our new results are probabilistic.
The number of edges used in all
constructions is $O(n)$. To save space, the $O$ notation is omitted
everywhere except for the exponents. The letters $\delta$ and
$\zeta$ stand for arbitrarily small positive constants.} 
\end{table}
}

\subsection{Background and Notation} \label{backnot}
In what follows, let $s > 0$ be a real fixed number.

We say that two point sets in the plane $A$ and $B$ are 
\emph{well-separated} with respect to $s$ if $A$ and $B$ can be
enclosed in two circles of radius $r$,
such that the distance between the two circles is at least $s \cdot r$.
The number $s$ is called the \emph{separation ratio} of $A$ and $B$.
A \emph{well-separated pair decomposition} (WSPD) for a point set $P$
in the plane with respect to $s$ is a set $\{\{A_1,B_1\},\{A_2,B_2\},\ldots,\{A_m,B_m\}\}$
of pairs of nonempty subsets of $P$, for some integer $m$, such that:
1) For each $i \in [m]$, $A_i$ and $B_i$ are well-separated with respect to $s$,
2) For any two distinct points $p$ and $q$ of $P$, there is exactly one index $i$
in $[m]$, such that either $p \in A_i$ and $q \in B_i$, or $p \in B_i$ and $q \in A_i$.

Next, we describe a well-known algorithm due to Callahan and Kosaraju \cite{CK92} for computing a WSPD for $P$ with respect to $s$.
The algorithm consists of two phases. In the first phase, we construct a \emph{split tree}, that is, a tree that corresponds to
a hierarchical
decomposition of $P$ into rectangles of bounded aspect ratio, where
rectangles serve as vertices of the tree, each being split into smaller
rectangles as long as it contains more than one point of $P$.
%
Observe that the split tree does not depend on $s$.
In the second phase, we employ the split tree to construct the WSPD itself.

There are many variants of a split tree, and
we outline below the fair split tree due to Callahan and Kosaraju \cite{CK92}.
Place a smallest-possible rectangle $R(P)$ about the point set $P$. The root of the fair
split tree is $R(P)$. Choose the longer side of $R(P)$ and divide it into two equal parts,
thus splitting $R(P)$ into two smaller rectangles of equal size, $R_l$ and $R_r$. The
left and right subtrees of the root $R(P)$ are the fair split trees that are constructed recursively
for the point sets $R_l \cap P$ and $R_r \cap P$, respectively. This recursive process is repeated until
a single point remains, in which case the split tree consists of just a single vertex that
stores this point.
Following Arya et al.\ \cite{ADMSS95},
we consider a fair split tree in an ideal form, 
henceforth the \emph{idealized box split tree}. In
this tree rectangles are squares,
each split recursively into four identical squares of half the 
side length. In other words, the idealized box split tree is a \emph{quadtree}. (Refer to
Chapter 14 of \cite{BCKO08} for the definition of quadtree.)
While actual constructions will be performed using  
the fair split tree or other closely related variants (see, e.g., 
the \emph{compressed quadtrees} of \cite{FHP05} and \cite{Chan08}, 
and the \emph{balanced box-decomposition tree} of \cite{AMNSW98}), 
the idealized
box split tree provides a clean and elegant way of conceptualizing 
the fair split tree in all its variants for purposes of analysis. 

\begin{figure*}[htp]
\begin{center}
\begin{minipage}{\textwidth} 
\begin{center}
\setlength{\epsfxsize}{2.2in} \epsfbox{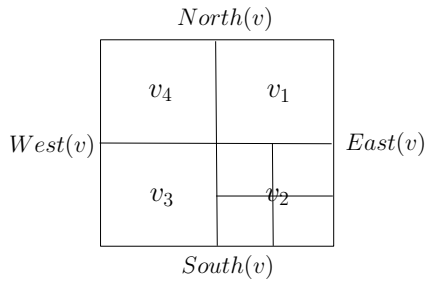}
\end{center}
\end{minipage}
\caption[]{ \label{square} \sf \footnotesize An illustration of a typical internal vertex $v$ in $\mathcal T$.
The vertex $v$ has four children $v_1,v_2,v_3$ and $v_4$, each being a square of half the side length $side(v)/2$.
Each child $v_i$ of $v$, $i \in [4]$, has four children of its own (unless it is a leaf), of side length $side(v)/2^2$ each, and so on. 
In the illustration only the four children of $v_2$ are depicted.}
\end{center}
\end{figure*}

Consider the idealized box split tree $\mathcal T = \mathcal T(P)$ that is constructed for $P$.
We identify each vertex $v$ in the tree $T$ with the square in the plane corresponding to it. 
For example, the root $rt = rt(\mathcal T)$ of $\mathcal T$ is identified with the smallest-possible square $R(P)$ 
about the point set $P$. Thus referring to, e.g., the side length of a vertex $v$ in $\mathcal T$,
is well-defined.
Suppose without loss of generality that the sides of the square $rt$ are parallel
to the $x$ and $y$ axes. 
Consequently, each vertex $v$ of $\mathcal T$ is a square
whose sides are parallel to the $x$ and $y$ axes. Denote the four sides of $v$ by $North(v)$, $South(v)$, $East(v)$ and
$West(v)$, with $North(v)$ and $South(v)$ (respectively, $East(v)$ and $West(v)$)
being parallel to the $x$-axis (resp., $y$-axis). 
Denote the four children of an internal vertex $v$ in $\mathcal T$ by $v_1,v_2,v_3$ and $v_4$, each being a square of half the side length
of $v$, where $v_1$, $v_2$, $v_3$ and $v_4$ are the North-Eastern, South-Eastern, South-Western and North-Western parts of $v$,
respectively. The side length of $v$ is denoted by $side(v)$. Notice that 
each child of $rt$ has side length
$\frac{side(rt)}{2}$, each grandchild of $rt$ has side length $\frac{side(rt)}{2^2}$, etc. More generally,
a vertex $v$ in $\mathcal T$ of level\footnote{The \emph{level} of a vertex in a rooted tree is defined as its unweighted distance from the root.} $L(v)$, $0 \le L(v) \le depth(\mathcal T)$, has side length $side(v) = \frac{side(rt)}{2^{L(v)}}$.
(See Figure \ref{square} for an illustration.)
Define $P(v) = v \cap P$. The vertex $v$ is called \emph{empty} if $P(v) = \emptyset$. 
Otherwise, it is \emph{non-empty}.
The depth of a vertex $v$ in $\mathcal T$ is defined as the
depth of the subtree $\mathcal T_v$ of $\mathcal T$ rooted at $v$.
For any two vertices $u$ and $v$ in $\mathcal T$, we denote by $dist(u,v)$  
the distance of closest approach between $u$ and $v$, i.e., 
the minimum distance between a point lying on the boundary of $u$ and a point lying
on the boundary of $v$.
Also, we denote by $distMax(P(u),P(v))$ the maximum 
distance between a point in $P(u)$ and a point in $P(v)$. Clearly, $distMax(P(u),P(v))$ is no smaller
than $dist(u,v)$. On the other hand,
it is bounded from above by the distance
of furthest approach between $u$ and $v$, i.e., the maximum distance
between a point lying on the boundary of $u$ and a point lying on the boundary of $v$,
which is, in turn, bounded from above by $dist(u,v) + 2\sqrt{2} \cdot
\max\{side(u),side(v)\}$. Thus, $dist(u,v) \le distMax(P(u),P(v)) \le 
dist(u,v) + 2\sqrt{2} \cdot
\max\{side(u),side(v)\}$.


To compute the WSPD of $P$, we use a simple recursive algorithm 
which consists of the two procedures below.
(This algorithm is essentially taken from Callahan and Kosaraju \cite{CK92}.)
%
We initially invoke Procedure 1 below by making the call $WSPD(rt(\mathcal T))$, where 
$rt(\mathcal T) = R(P)$. The output returned by this call is the WSPD for $P$. We omit the proof
of correctness, which resembles that of \cite{CK92}.     
Notice that for any pair of vertices $u$ and $v$ in  $\mathcal T$, both calls
$WSPD(u,v)$ and $WSPD(u)$ return sets of well-separated pairs of $P$. In what follows we
write $WSPD(u,v)$ and $WSPD(u)$ to refer to the sets that are returned by these calls 
(rather than to the calls themselves).
%
\\$\line(1,0){500}$
\\{\bf Procedure 1}~~~$WPSD(u):$
\begin{algorithmic}[1]
  \IF {$|P(u)| \le 1$} 
  \STATE return $\emptyset$
  \ENDIF
  \STATE return $\bigcup_{1 \le i \le 4} WSPD(u_i) \cup \bigcup_{1 \le i < j \le 4} WSPD(u_i,u_j)$
\end{algorithmic}
\vspace{-0.1in}
$\line(1,0){500}$
\\$\line(1,0){500}$
{\bf Procedure 2}~~~$WPSD(u,v):$
\begin{algorithmic}[1]
  \IF {$P(u) = \emptyset$ or $P(v) = \emptyset$} 
  \STATE return $\emptyset$
  \ENDIF
  \IF {$P(u)$ and $P(v)$ are well-separated} 
  \STATE return $\{\{P(u),P(v)\}\}$
  \ENDIF
  \IF {$side(u) \ge side(v)$}
  \STATE return $\bigcup_{1 \le i \le 4} WSPD(u_i,v)$
  \ELSE 
  \STATE return $\bigcup_{1 \le i \le 4} WSPD(u,v_i)$
  \ENDIF
\end{algorithmic}
\vspace{-0.12in}
$\line(1,0){500}$


\vspace{0.1in}
A \emph{representative assignment} for the split tree $\mathcal T= \mathcal T(P)$
is a mapping $\varphi$ between vertices of $\mathcal T$ and points of $P$,
sending each vertex $v$ in $\mathcal T$ to a point $\varphi(v)$ in $P(v)$.
%
The point $\varphi(v)$ is called the 
\emph{representative} of $v$ under the mapping $\varphi$.
We say that a pair $(A,B)$ of nonempty sets of $P$ \emph{belongs} to $\mathcal T$, if there
are two vertices $u$ and $v$ in $\mathcal T$, such that $A = P(u)$ and $B = P(v)$.
Given a representative assignment $\varphi$, there is a natural correspondence between
a well-separated pair $\{P(u),P(v)\}$ that 
belongs to $\mathcal T$ and the edge 
$(\varphi(u),\varphi(v))$ connecting the representatives of $u$ and $v$ under $\varphi$. 
In the same way, there is a natural correspondence between a set $S$ of well-separated pairs of $P$ that belong to $\mathcal T$ 
and the edge set 
 $E_{\varphi}(S)$, where $E_{\varphi}(S) = \{(\varphi(u),\varphi(v)) ~\vert~ 
\{P(u),P(v)\} \in S\}$. 
The weight $w(H)$ of an edge set $H$ is defined as the sum $\sum_{e=(u,v) \in H} w(u,v)$ of all edge weights in it,
where $w(u,v) = \|u-v\|$.
Callahan and Kosaraju \cite{CK93} showed that for any
representative assignment $\varphi$, the edge set $E^* = E_{\varphi}(WSPD(P))$ that corresponds to  
$WSPD(P) = WSPD(rt(\mathcal T))$ constitutes a $(1+\eps)$-spanner (with $O(n)$ edges), henceforth the \emph{WSPD-spanner} of $P$,
where $\eps$ is an arbitrarily small constant depending on $s$. (It can be easily shown that $\eps \le \frac{8}{s-4}$.)

\subsection{Proof of Proposition \ref{maini}} \label{arg}
In this section we prove Proposition \ref{maini}.

Let $P$ be an arbitrary set of $n$ points in the plane, and
let $\mathcal T = \mathcal T(P)$ and $WSPD(P) = WSPD(rt(\mathcal T))$ be the idealized box split tree 
and the WSPD that are constructed for it, respectively.
Also, fix an arbitrary representative assignment $\varphi$ for $\mathcal T$.
Next, we show that the weight $w(E^*)$ of the WSPD-spanner $E^* = E_{\varphi}(WSPD(P))$ is 
at most $c^* \cdot side(rt(\mathcal T)) \cdot \sqrt{n}$,
where $c^*$ is a sufficiently large constant that depends only on $s$. 
(We do not try to optimize the constant $c^*$.)
In particular, for a point set $P$ in the unit square we have $side(rt(\mathcal T))=1$,
thus proving Proposition \ref{maini}.

Observe that for any two vertices $u$ and $v$ in $\mathcal T$, both $WSPD(u,v)$ and $WSPD(u)$
are sets of well-separated pairs of $P$ that belong to $\mathcal T$. 
Henceforth, we write $W(u,v)$ and $W(u)$ as shortcuts for
$w(E_{\varphi}(WSPD(u,v)))$ and $w(E_{\varphi}(WSPD(u)))$, respectively. 

\begin{lemma} \label{packing}
Let $u$ and $v$ be two vertices in $\mathcal T$, such that 
$dist(u,v) = c \cdot \max\{side(u),side(v)\}$, for some constant $1/2 \le c \le \sqrt{2}$.
Then $W(u,v) \le \alpha \cdot \max\{side(u),side(v)\}$, where $\alpha = \alpha_s$ is a sufficiently large 
constant that depends only on $s$.
\end{lemma}
\begin{proof}
From standard packing arguments, it follows that $|WSPD(u,v)| \le \tilde \alpha$, where $\tilde \alpha = \tilde \alpha_s$ is a 
sufficiently large constant that depends
only on $s$.
For each pair $\{P(x),P(y)\}$ in $WSPD(u,v)$, the weight $\|\varphi(x)-\varphi(y)\|$ of the
corresponding edge $(\varphi(x),\varphi(y))$ is at most $dist(u,v) + 2 \sqrt{2} \cdot \max\{side(u),side(v)\} ~=~\\
(c + 2\sqrt{2}) \cdot \max\{side(u),side(v)\}$.
Define $\alpha = \tilde \alpha (c + 2\sqrt{2})$.
It follows that $W(u,v) ~\le~ \tilde \alpha (c + 2\sqrt{2}) \cdot \max\{side(u),side(v)\} ~=~ \alpha \cdot  \max\{side(u),side(v)\}.$ \qed
\end{proof}

\begin{figure*}[htp]
\begin{center}
\begin{minipage}{\textwidth} 
\begin{center}
\setlength{\epsfxsize}{3.3in} \epsfbox{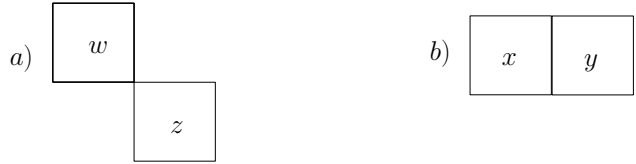}
\end{center}
\end{minipage}
\caption[]{ \label{diagadj1} \sf \footnotesize a) Two diagonal vertices $w$ and $z$.
b) Two adjacent vertices $x$ and $y$.} 
\end{center}
\end{figure*}

\begin{figure*}[htp]
\begin{center}
\begin{minipage}{\textwidth} 
\begin{center}
\setlength{\epsfxsize}{4.7in} \epsfbox{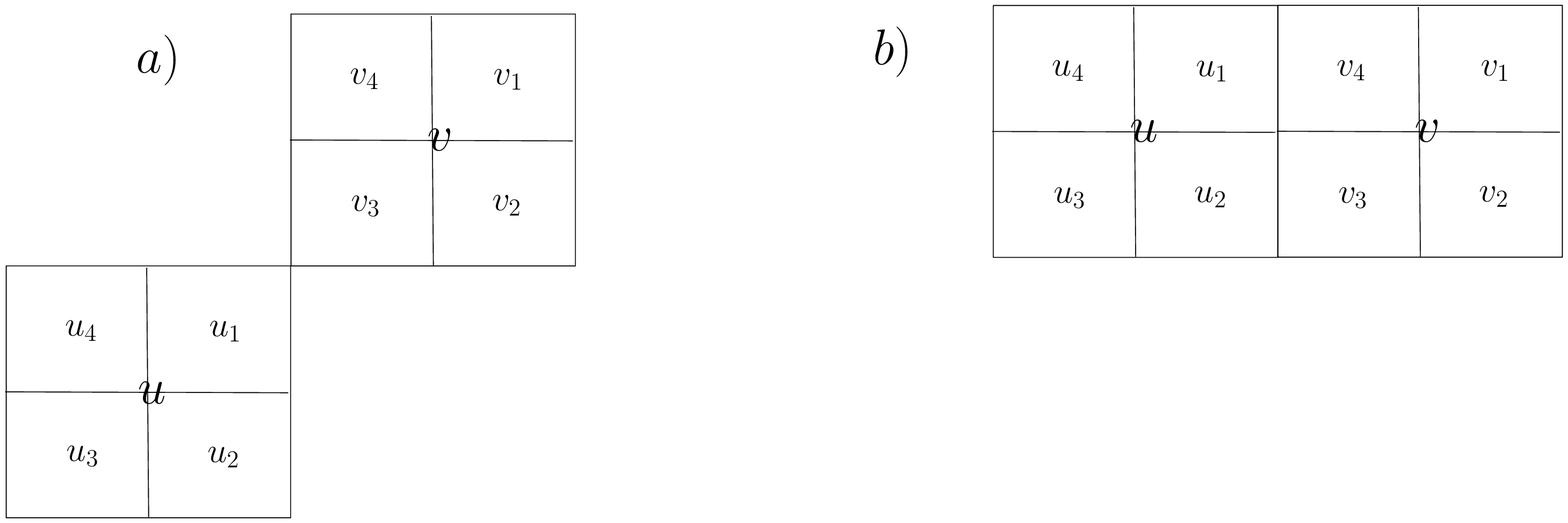}
\end{center}
\end{minipage}
\caption[]{ \label{diagadj} \sf \footnotesize 
a) An illustration of two diagonal vertices $u$ and $v$. 
The vertex $u_1$ is diagonal to both
$u_3$ and $v_3$. The vertices $u_1$ and $v_3$ intersect at the same point
as their respective parents $u$ and $v$ do. b)
An illustration of two adjacent vertices $u$ and $v$. 
The vertex $u_1$ is adjacent to $u_2$,
$u_4$ and $v_4$. 
The vertices $u_1$ and $v_4$ intersect at a single side,
which is the upper half of the side at which their respective parents $u$ and $v$ intersect.}
\end{center}
\end{figure*}

We say that two vertices $u$ and $v$ in $\mathcal T$ of the same level are \emph{diagonal} if
their boundaries intersect at a single point. (See Figure \ref{diagadj1}.a for an illustration.)
For example, for any vertex $v$ in $\mathcal T$, its two children $v_1$ and $v_3$ are diagonal.  
Consider two diagonal vertices $u$ and $v$ in $\mathcal T$. 
Since by definition they are at the same level in $\mathcal T$, 
it holds that $side(u) = side(v)$.
Also, notice that
 $dist(u,v) = 0$ and $\|\varphi(u) - \varphi(v)\| \le distMax(P(u),P(v)) \le 2 \sqrt{2} \cdot side(u)$. 
\begin{lemma} \label{previous}
For any two diagonal vertices $u$ and $v$ in 
$\mathcal T$, $W(u,v) \le \beta \cdot side(u)$, where $\beta = \beta_s$ is a sufficiently large constant that depends
only on $s$.
\end{lemma}
\begin{proof}
The proof is by induction on the sum $h = depth(u) + depth(v)$ of depths of $u$ and $v$.
\\\emph{Basis: $h=0$.} In this case both $u$ and $v$ are leaves, and so each
one of them contains at most one point.
If either $u$ or $v$ is empty, then $WSPD(u,v)$ is an empty set, and so $W(u,v) = 0 < \beta \cdot side(u)$.
Otherwise, $WSPD(u,v) = \{\{P(u),P(v)\}\}$, and so $W(u,v) = \|\varphi(u)-\varphi(v)\| \le 2 \sqrt{2} \cdot side(u)
< \beta \cdot side(u)$. 
\\\emph{Induction Step:} We assume the correctness of the statement for all smaller values of $h$, $h \ge 1$, and prove it for $h$.
If either $u$ or $v$ is empty, then $WSPD(u,v)$ is an empty set, 
and so $W(u,v) = 0 < \beta \cdot side(u)$. Otherwise, if $P(u)$ and $P(v)$ are well-separated
then $WSPD(u,v) = \{\{P(u),P(v)\}\}$, and so $W(u,v)  = \|\varphi(u) - \varphi(v)\| \le 2 \sqrt{2} \cdot side(u) <
\beta \cdot side(u)$. We henceforth assume that $P(u)$ and $P(v)$ are not well-separated. In this case
$WSPD(u,v) = \bigcup_{1 \le i \le 4} WSPD(u_i,v)$, and so $W(u,v) = \sum_{1 \le i \le 4} W(u_i,v)$. Since $u$ and $v$ are diagonal, the intersection of $v$ and exactly one child of $u$
consists of a single point, whereas all the other children of $u$ are disjoint from $v$.
Suppose without loss of generality that  the child of $u$ that intersects $v$ is $u_1$. 
(See Figure \ref{diagadj}.a for an illustration.)
Observe that
$dist(u_2,v) = dist(u_4,v) = \frac{1}{2} \cdot side(v)$ and 
$dist(u_3,v) = \frac{1}{\sqrt{2}} \cdot side(v)$.
Hence, by Lemma \ref{packing}, 
for each $2 \le i \le 4$, 
$W(u_i,v) \le \alpha \cdot side(v)$. Next, we bound $W(u_1,v)$.
If $u_1$ is empty, then $WSPD(u_1,v)$ is an empty set, and so  $W(u_1,v) = 0$. Also, if $P(u_1)$
and $P(v)$  are well-separated, then
$WSPD(u_1,v)
= \{\{P(u_1),P(v)\}\}$, and so $W(u_1,v)
= \|\varphi(u_1)-\varphi(v)\| \le 2 \sqrt{2} \cdot side(v)$. Otherwise, $WSPD(u_1,v) = 
\bigcup_{1 \le i \le 4} WSPD(u_1,v_i)$, and so
$W(u_1,v) = 
\sum_{1 \le i \le 4} W(u_1,v_i)$. Observe that 
$dist(u_1,v_2) = dist(u_1,v_4) = side(u_1)$ and $dist(u_1,v_1) = \sqrt{2} \cdot side(u_1)$. 
Hence, by Lemma \ref{packing}, for each $i \ne 3$, $W(u_1,v_i) \le \alpha \cdot side(u_1) = \frac{\alpha}{2} \cdot side(u)$.
Notice that $u_1$ and $v_3$ are diagonal, and so
by the induction hypothesis, $W(u_1,v_3) \le \beta \cdot side(u_1) = \frac{\beta}{2} \cdot side(u)$.
Set $\beta = 9\alpha$. Altogether, 
\begin{eqnarray*}
W(u,v) &=&  
\sum_{2 \le i \le 4} W(u_i,v) + W(u_1,v) ~\le~  
3\alpha \cdot side(v) + \sum_{1 \le i \le 4, i \ne 3}W(u_1,v_i) + W(u_1,v_3)
\\ &\le& 3\alpha \cdot side(v) + 3 \cdot \frac{\alpha}{2} \cdot side(u) + \frac{\beta}{2} \cdot side(u) ~=~
side(u) \cdot \left(\frac{9\alpha}{2}  + \frac{\beta}{2}\right) ~=~ \beta \cdot side(u). 
\end{eqnarray*}
\qed
\end{proof}

We say that two vertices $u$ and $v$ in $\mathcal T$ of the same level are \emph{adjacent} if
their boundaries intersect at a single side. (See Figure \ref{diagadj1}.b for an illustration.) 
For example, for any vertex $v$ in $\mathcal T$, its two children $v_1$ and $v_2$ are adjacent.
Consider two adjacent vertices $u$ and $v$ in $\mathcal T$. 
Since by definition they are at the same level in $\mathcal T$, 
it holds that $side(u) = side(v)$.
Also, notice that $dist(u,v) = 0$ and
$\|\varphi(u) - \varphi(v)\| \le distMax(P(u),P(v)) \le \sqrt{5} \cdot side(u)$.
For a vertex $v$ in $\mathcal T$, define $N(v) = |P(v)|$. 
\begin{lemma} \label{adji}
For any two adjacent vertices $u$ and $v$ in 
$\mathcal T$ such that $N(u) + N(v) \ge 1$, 
$W(u,v) \le \gamma \cdot side(u) \cdot \log(N(u)+N(v))$, where $\gamma = \gamma_s$ is a sufficiently large constant that depends
only on $s$.
\end{lemma}
\begin{proof}
The proof is by induction on the sum $h = depth(u) + depth(v)$ of depths of $u$ and $v$.
\\\emph{Basis: $h=0$.} In this case both $u$ and $v$ are leaves, and so each
one of them contains at most one point.
If either $u$ or $v$ is empty, then $WSPD(u,v)$ is an empty set, and so $W(u,v) = 0  = \gamma \cdot side(u) \cdot \log 1$.
Otherwise, $WSPD(u,v) = \{\{P(u),P(v)\}\}$, and so $W(u,v) = \|\varphi(u)-\varphi(v)\| \le \sqrt{5} \cdot side(u)
< \gamma \cdot side(u) \cdot \log 2$. 
\\\emph{Induction Step:} We assume the correctness of the statement for all smaller values of $h$, $h \ge 1$, and prove it for $h$.
If either $u$ or $v$ is empty, then $WSPD(u,v)$ is an empty set, 
and so $W(u,v) = 0 \le \gamma \cdot side(u) \cdot \log (N(u)+N(v))$. Otherwise, if $P(u)$ and $P(v)$ are well-separated
then $WSPD(u,v) = \{\{P(u),P(v)\}\}$, and so $W(u,v)  = \|\varphi(u) - \varphi(v)\| \le \sqrt{5} \cdot side(u) <
\gamma \cdot side(u) \cdot \log (N(u)+N(v))$. We henceforth assume that $P(u)$ and $P(v)$ are not well-separated. In this case
$WSPD(u,v) = \bigcup_{1 \le i \le 4} WSPD(u_i,v)$, and so $W(u,v) = \sum_{1 \le i \le 4} W(u_i,v)$. 
Since $u$ and $v$ are adjacent, exactly two adjacent children $u_i$ and $u_{i+1}$
of $u$ intersect $v$, $i \in [3]$, each at a single side.
Suppose without loss of generality that these children of $u$ are $u_1$ and $u_2$. 
(See Figure \ref{diagadj}.b for an illustration.)
Observe that
$dist(u_3,v) = dist(u_4,v) = \frac{1}{2} \cdot side(v)$.
Hence, by Lemma \ref{packing},
$W(u_3,v),W(u_4,v) \le \alpha \cdot side(v)$. Next, we bound $W(u_1,v)$.     
If $u_1$ is empty, then $WSPD(u_1,v)$ is an empty set, and so $W(u_1,v) = 0$. 
Also, if  $P(u_1)$ and $P(v)$ are well-separated, then $WSPD(u_1,v) = \{\{P(u_1),P(v)\}\}$, and so $W(u_1,v)
= \|\varphi(u_1)-\varphi(v)\| \le \sqrt{5} \cdot side(v)$. Otherwise, $WSPD(u_1,v) = 
\bigcup_{1 \le i \le 4} WSPD(u_1,v_i)$, and so $W(u_1,v) = 
\sum_{1 \le i \le 4} W(u_1,v_i)$. Observe that $dist(u_1,v_1) = dist(u_1,v_2) = side(u_1)$. 
Hence, by Lemma \ref{packing}, $W(u_1,v_1),W(u_1,v_2) \le \alpha \cdot side(u_1) = \frac{\alpha}{2} \cdot side(u)$.
Notice that $u_1$ and $v_3$ are diagonal, 
whereas $u_1$ and $v_4$ are adjacent. 
Hence, by Lemma \ref{previous}, $W(u_1,v_3) \le \beta \cdot side(u_1) = \frac{\beta}{2} \cdot side(u)$.
Recall that $u_1$ is non-empty, and so $N({u_1}) + N({v_4}) \ge 1$.
By the induction hypothesis, $W(u_1,v_4) \le \gamma \cdot side(u_1) \cdot \log(N({u_1}) + N({v_4})) = \frac{\gamma}{2} \cdot side(u) \cdot \log(N({u_1}) + N({v_4}))$. We get that \begin{eqnarray*}
W(u_1,v) &=& \sum_{1 \le i \le 4} W(u_1,v_i) 
~\le~ \alpha \cdot side(u) + \frac{\beta}{2} \cdot side(u) + \frac{\gamma}{2} \cdot side(u) \cdot \log(N({u_1}) + N({v_4}))
\\ &=& side(u) \cdot \left(\alpha + \frac{\beta}{2} + \frac{\gamma}{2} \cdot \log(N({u_1}) + N({v_4}))\right).
\end{eqnarray*}
A symmetric argument yields $W(u_2,v) ~\le~ side(u) \cdot \left(\alpha + \frac{\beta}{2} + \frac{\gamma}{2} \cdot \log(N({u_2}) + N({v_3}))\right).$
\\Observe that $N({u_1}) + N({u_2}) \le N(u)$ and $N({v_3}) + N({v_4}) \le N(v)$, implying that
$2(N({u_1}) + N({v_4}))\cdot(N({u_2}) + N({v_3})) \le (N(u)+N(v))^2$.
Set $\gamma = 2(4\alpha + \beta)$. 
Altogether, 
\begin{eqnarray*} \label{secondcase2}
\nonumber W(u,v) &=& \sum_{1 \le i \le 4} W(u_i,v) ~=~
 [W(u_1,v) + W(u_2,v)] + [W(u_3,v) + W(u_4,v)] \\ \nonumber &\le& 
 \left[side(u) \cdot \left(2\alpha + {\beta} + \frac{\gamma}{2} \cdot \left(\log(N({u_1}) + N({v_4})) + \log(N({u_2}) + N({v_3}))\right)\right)\right] 
+ \left[2\alpha \cdot side(v)\right] \\ &=& \gamma \cdot side(u)  \cdot \left(\frac{4\alpha}{\gamma} + \frac{\beta}{\gamma} + \frac{1}{2} \cdot \left(\log(N({u_1}) + N({v_4})) + \log(N({u_2}) + N({v_3}))\right)\right) \\ &=& \gamma \cdot side(u) \cdot \left(\frac{1+\log(N({u_1}) + N({v_4})) + \log(N({u_2}) + N({v_3}))}{2}\right) \\ &=& \gamma \cdot side(u) \cdot \log \sqrt{2(N({u_1}) + N({v_4}))\cdot(N({u_2}) + N({v_3}))}
~\le~  \gamma \cdot side(u)  \cdot \log (N(u)+N(v)).
\end{eqnarray*}
\qed
\end{proof}

We use the following claim to prove Lemma \ref{meaning2}.
\begin{claim} \label{techbound}
For any positive integers $n_1,n_2,\ldots,n_k$, $k$ and $n$, such that $\sum_{i=1}^k n_i = n$, 
$$\sum_{i=1}^k \left(\sqrt{n_i} - \frac{\ln n_i}{8}\right) ~\le~ f(n,k) ~=~ k \cdot \left(\sqrt{n/k} - \frac{\ln(n/k)}{8}\right).$$
\end{claim}
\begin{proof}
The proof is by induction on $k$, for $k \in [n]$. The basis $k=1$ is trivial. 
\\\emph{Induction Step:}
We assume the correctness of the statement for all smaller values of
$k$, $k \ge 2$, and prove it for $k$. 
By the induction hypothesis, 
$$\sum_{i=1}^{k-1} \left(\sqrt{n_i} - \frac{\ln n_i}{8}\right) ~\le~ f(n-n_k,k-1) ~=~ (k-1) \cdot \left(\sqrt{\frac{n-n_k}{k-1}} - \frac{\ln\left(\frac{n-n_k}{k-1}\right)}{8}\right).$$
It follows that \begin{equation} \label{stamla}
\sum_{i=1}^k \left(\sqrt{n_i} - \frac{\ln n_i}{8}\right) ~\le~ (k-1) \cdot \left(\sqrt{\frac{n-n_k}{k-1}} - \frac{\ln\left(\frac{n-n_k}{k-1}\right)}{8}\right) + \sqrt{n_k} - \frac{\ln n_k}{8}.
\end{equation}
Define $g_{n,k}(x) = (k-1) \cdot \left(\sqrt{\frac{n-x}{k-1}} - \frac{\ln\left(\frac{n-x}{k-1}\right)}{8}\right) + \sqrt{x} - \frac{\ln x}{8}$. Since $n_1,n_2,\ldots,n_k \ge 1$ are positive integers and $\sum_{i=1}^k n_i = n$, we have that $1 \le n_k \le n-k+1$. Hence, the maximum value of the function $g_{n,k}(x)$ in the range $1 \le x \le n-k+1$ provides an
upper bound on the right-hand side of (\ref{stamla}). 
It is easy to verify that the function $g_{n,k}(x)$ in the range $1 \le x \le n-k+1$ is maximized at $x = n/k$. 
Hence, in the range $1 \le x \le n-k+1$, $g_{n,k}(x) \le g_{n,k}(n/k) = k \cdot \left(\sqrt{n/k} - \frac{\ln(n/k)}{8}\right)$,
and we are done.
\qed
\end{proof}

The next lemma implies that $w(E^*) = W(rt) \le 
c^* \cdot side(rt) \cdot \left(\sqrt{n} - \frac{\ln (n)}{8} \right)
\le c^* \cdot side(rt) \cdot \sqrt{n}$. 
Hence,
for any set of $n$ points in the unit square,
the weight of the WSPD-spanner is $O(\sqrt{n})$, thus proving Proposition \ref{maini}. 
\begin{lemma} \label{meaning2}
For any non-empty vertex $u$ in 
$\mathcal T$, 
$W(u) \le c^* \cdot side(u) \cdot \left(\sqrt{N(u)} - \frac{\ln (N(u))}{8} \right)$.
\end{lemma}
\begin{proof}
The proof is by induction on the depth $h= depth(u)$ of $u$. The basis $h=0$ is trivial.
\\\emph{Induction Step:} We assume the correctness of the statement for all smaller values of $h$, $h \ge 1$, and prove it for $h$.
First, suppose that $1 \le N(u) < 20$. In this case, we have $|WSPD(u)| \le c$,
for a sufficiently large constant $c$. Also, the weight of every edge in the edge set that corresponds to $WSPD(u)$ is 
at most $\sqrt{2} \cdot side(u)$, and so $$W(u) \le c \cdot \sqrt{2} \cdot side(u) < c^* \cdot side(u) \cdot \left(\sqrt{N(u)} - \frac{\ln (N(u))}{8} \right).$$
We henceforth assume that $N(u) \ge 20$. Hence, 
$$WSPD(u) = \bigcup_{1 \le i \le 4} WSPD(u_i) \cup  \bigcup_{1 \le i < j \le 4} WSPD(u_i,u_j),$$ and so
\begin{equation} \label{dayan} 
W(u) = \sum_{1 \le i \le 4} W(u_i) + \sum_{1 \le i < j \le 4} W(u_i,u_j).
\end{equation}
To bound $W(u)$, we start with bounding the left sum $\sum_{1 \le i \le 4} W(u_i)$
in the right-hand side of (\ref{dayan}).
Denote by $I$ the set of indices in $[4]$ for which $N({u_i}) \ge 1$. 
By the induction hypothesis, for each index $i \in I$,
$W(u_i) \le  c^* \cdot side(u_i) \cdot \left(\sqrt{N({u_i})} - \frac{\ln (N({u_i}))}{8} \right)$.
Also, for each index $i \in [4] \setminus I$, we have $W(u_i) = 0$.
It follows that \begin {eqnarray} \label{jirk}
\nonumber \sum_{1 \le i \le 4} W(u_i) &=& \sum_{i \in I} W(u_i) ~\le~ \sum_{i \in I}
c^* \cdot side(u_i) \cdot \left(\sqrt{N({u_i})} - \frac{\ln (N({u_i}))}{8} \right)
\\ &=&  \frac{c^*}{2} \cdot side(u) \cdot \sum_{i \in I} \left(\sqrt{N({u_i})} - \frac{\ln (N({u_i}))}{8}\right).
\end{eqnarray}
Observe that $\sum_{i \in I} N({u_i}) = N(u)$ and $1 \le |I| \le 4$. 
By Claim \ref{techbound},
$$\sum_{i \in I} \left(\sqrt{N({u_i})} - \frac{\ln (N({u_i}))}{8}\right) ~\le~ f(N(u),|I|) ~=~ |I| \cdot \left(\sqrt{N(u)/|I|} - \frac{\ln(N(u)/|I|)}{8}\right).
$$
It is easy to verify that the function $f_{N(u)}(x) = f(N(u),x) = x \cdot \left(\sqrt{N(u)/x} - \frac{\ln(N(u)/x)}{8}\right)$ 
is monotone increasing with $x$ 
in the range $x > 0$.
(The derivative $f'_{N(u)}(x)$ is strictly positive for all $x > 0$.)
Since $|I| \le 4$, we thus have 
\begin{eqnarray} \label{tensharpit}
\nonumber \sum_{i \in I} \left(\sqrt{N({u_i})} - \frac{\ln (N({u_i}))}{8}\right) &\le& f(N(u),|I|) ~\le~ f(N(u),4) ~=~ 4 \cdot \left(\sqrt{N(u)/4} - \frac{\ln(N(u)/4)}{8}\right)  \\ &=& 2 \cdot \sqrt{N(u)} - \frac{\ln(N(u)/4)}{2}.
\end{eqnarray}
Plugging (\ref{tensharpit}) into (\ref{jirk}) yields
\begin{equation} \label{gigi}
\sum_{1 \le i \le 4} W(u_i) ~\le~ \frac{c^*}{2} \cdot side(u) \cdot \left(2 \cdot \sqrt{N(u)} - \frac{\ln(N(u)/4)}{2}\right)
~=~ c^* \cdot side(u) \cdot \left(\sqrt{N(u)} - \frac{\ln(N(u)/4)}{4}\right).
\end{equation}
We proceed with bounding the right sum  $\sum_{1 \le i < j \le 4} W(u_i,u_j)$ in the right-hand side of (\ref{dayan}).
Observe that the two pairs $(u_1,u_3)$ and $(u_2,u_4)$ of children of $u$ are diagonal,
whereas the four other pairs $(u_1,u_2), (u_1,u_4), (u_2,u_3)$ and $(u_3,u_4)$ are adjacent.
By Lemma \ref{previous}, $W(u_1,u_3), W(u_2,u_4) \le \beta \cdot side(u_1) = \frac{\beta}{2} \cdot side(u)$.
Consider a pair $(u_i,u_j)$ among the four pairs of adjacent children of $u$.
If both $u_i$ and $u_j$ are empty, then $W(u_i,u_j) = 0$.
Otherwise, we have $N({u_i}) + N({u_j}) \ge 1$, and so 
by Lemma \ref{adji},
\begin{eqnarray*}
W(u_i,u_j) ~\le~ \gamma \cdot side(u_i) \cdot \log(N({u_i})+N({u_j})) ~\le~ \gamma \cdot side(u_i) \cdot \log(N(u))
~=~ \frac{\gamma}{2} \cdot side(u) \cdot \log(N(u)).\end{eqnarray*}
Recall that $\gamma = 2(4\alpha + \beta)$, and so $\beta \le \frac{\gamma}{2}$. 
Altogether \begin{eqnarray} \label{backhurts}
\nonumber \sum_{1 \le i < j \le 4} W(u_i,u_j) &\le& \beta \cdot side(u) + 2\gamma \cdot side(u) \cdot \log(N(u))
\\ &\le& \gamma \cdot side(u) \cdot \left(\frac{1}{2} + 2\log(N(u))\right) 
 ~\le~ 4\gamma \cdot side(u) \cdot \ln(N(u)).
\end{eqnarray}
Plugging (\ref{gigi}) and (\ref{backhurts}) into (\ref{dayan}) yields
\begin{eqnarray} \label{di}
\nonumber W(u) &=& \sum_{1 \le i \le 4} W(u_i) + \sum_{1 \le i < j \le 4} W(u_i,u_j)
\\&\le& 
c^* \cdot side(u) \cdot \left(\sqrt{N(u)} - \frac{\ln(N(u)/4)}{4}\right)
+ 4\gamma \cdot side(u) \cdot \ln(N(u)) .
\end{eqnarray}
It is easy to verify that for a sufficiently large constant $c^*$ and all $n \ge 20$, the
right-hand side of (\ref{di}) is no greater than 
$c^* \cdot side(u) \cdot \left(\sqrt{N(u)} - \frac{\ln (N(u))}{8} \right)$,
and we are done.
\qed
\end{proof}

\end{document}

%% file: latexmac.tex
\def\denseformat{
\setlength{\textheight}{9.5in}
\setlength{\textwidth}{6.9in}
\setlength{\evensidemargin}{-0.3in}
\setlength{\oddsidemargin}{-0.3in}
\setlength{\headsep}{10pt}
\setlength{\topmargin}{-0.44in}
\setlength{\columnsep}{0.375in}
\setlength{\itemsep}{0pt}
}

\newtheorem{theorem}{Theorem}[section]
\newtheorem{proposition}[theorem]{Proposition}

\newtheorem{claim}[theorem]{Claim}
\newtheorem{lemma}[theorem]{Lemma}

\newtheorem{corollary}[theorem]{Corollary}

\newtheorem{comment}[theorem]{Comment}

\def\boldhead#1:{\par\vskip 7pt\noindent{\bf #1:}\hskip 10pt}
\def\ithead#1:{\par\vskip 7pt\noindent{\it #1:}\hskip 10pt}

\def\inline#1:{\par\vskip 7pt\noindent{\bf #1:}\hskip 10pt}
\def\midinline#1:{\par\noindent{\bf #1:}\hskip 10pt}
\def\dnsinline#1:{\par\vskip -7pt\noindent{\bf #1:}\hskip 10pt}
\def\ddnsinline#1:{\newline{\bf #1:}\hskip 10pt}
\def\largeinline#1:{\par\vskip 7pt\noindent{\large\bf #1:}\hskip 10pt}
\long\def\comment #1\commentend{}
\long\def\commhide #1\commhideend{}
\long\def\commfull #1\commend{#1}
\long\def\commabs #1\commenda{}
\long\def\commtim #1\commendt{#1}
\long\def\commb #1\commbend{}
\long\def\commedit #1\commeditend{} 

\long\def\commB #1\commBend{}       

\long\def\commex #1\commexend{}     

\long\def\commsiena #1\commsienaend{}  
                                         
\long\def\commBI #1\commBIend{}  
                                         
\long\def\CProof #1\CQED{}
\def\qed{\mbox{}\hfill $\Box$\\}
\def\blackslug{\hbox{\hskip 1pt \vrule width 4pt height 8pt
    depth 1.5pt \hskip 1pt}}
\def\QED{\quad\blackslug\lower 8.5pt\null\par}

\def\Proof{\par\noindent{\bf Proof:~}}

\def\proof{\Proof}

\long\def\PPP#1{\noindent{\bf Proof:}{ #1}{\quad\blackslug\lower 8.5pt\null}}

\long\def\denspar #1\densend
{#1}

\newif\ifnotesw\noteswtrue
   {\ifnotesw\marginpar[\hfill\(\top\)]{\(\top\)}\fi}%
   {\ifnotesw\marginpar[\hfill\(\bot\)]{\(\bot\)}\fi}

\newcommand{\mnote}[1]%
    {\ifnotesw\marginpar%
        [{\scriptsize\it\begin{minipage}[t]{\marginparwidth}
        \raggedleft#1%
                        \end{minipage}}]%
        {\scriptsize\it\begin{minipage}[t]{\marginparwidth}
        \raggedright#1%
                        \end{minipage}}%
    \fi}

\def\tR{{\tilde R}}

\def\MathF{\hbox{\rm I\kern-2pt F}}
\def\MathP{\hbox{\rm I\kern-2pt P}}
\def\MathR{\hbox{\rm I\kern-2pt R}}
\def\MathZ{\hbox{\sf Z\kern-4pt Z}}
\def\MathN{\hbox{\rm I\kern-2pt I\kern-3.1pt N}}
\def\MathC{\hbox{\rm \kern0.7pt\raise0.8pt\hbox{\footnotesize I}
\kern-4.2pt C}}
\def\MathQ{\hbox{\rm I\kern-6pt Q}}

\newsavebox{\ttop}\newsavebox{\bbot}

\def\eps{\epsilon}




%% file: SubmissionCorr.bbl
\begin{thebibliography}{10}\setlength{\itemsep}{-1ex}\small

\bibitem{AWY05}
P.~K. Agarwal, Y.~Wang, and P.~Yin.
\newblock Lower bound for sparse {E}uclidean spanners.
\newblock In {\em Proc. of 16th SODA}, pages 670--671, 2005.

\bibitem{AS87}
N.~Alon and B.~Schieber.
\newblock Optimal preprocessing for answering on-line product queries.
\newblock {\em Manuscript}, 1987.

\bibitem{ADDJS93}
I.~Alth$\ddot{\mbox{o}}$fer, G.~Das, D.~P. Dobkin, D.~Joseph, and J.~Soares.
\newblock On sparse spanners of weighted graphs.
\newblock {\em Discrete \& Computational Geometry}, 9:81--100, 1993.

\bibitem{ADMSS95}
S.~Arya, G.~Das, D.~M. Mount, J.~S. Salowe, and M.~H.~M. Smid.
\newblock {E}uclidean spanners: short, thin, and lanky.
\newblock In {\em Proc. of 27th STOC}, pages 489--498, 1995.

\bibitem{AMNSW98}
S.~Arya, D.~M. Mount, N.~S. Netanyahu, R.~Silverman, and A.~Y. Wu.
\newblock An optimal algorithm for approximate nearest neighbor searching in
  fixed dimensions.
\newblock {\em J. ACM}, 45(6):819--923, 1998.

\bibitem{AS97}
S.~Arya and M.~H.~M. Smid.
\newblock Efficient construction of a bounded degree spanner with low weight.
\newblock {\em Algorithmica}, 17(1):33--54, 1997.

\bibitem{BGJRW09}
A.~Bhattacharyya, E.~Grigorescu, K.~Jung, S.~Raskhodnikova, and D.~P. Woodruff.
\newblock Transitive-closure spanners.
\newblock In {\em Proc. of 20th SODA}, pages 932--941, 2009.

\bibitem{BTS94}
H.~L. Bodlaender, G.~Tel, and N.~Santoro.
\newblock Trade-offs in non-reversing diameter.
\newblock {\em Nord. J. Comput.}, 1(1):111--134, 1994.

\bibitem{CK92}
P.~B. Callahan and S.~R. Kosaraju.
\newblock A decomposition of multi-dimensional point-sets with applications to
  $k$-nearest-neighbors and $n$-body potential fields.
\newblock In {\em Proc. of 24th STOC}, pages 546--556, 1992.

\bibitem{CK93}
P.~B. Callahan and S.~R. Kosaraju.
\newblock Faster algorithms for some geometric graph problems in higher
  dimensions.
\newblock In {\em Proc. of 4th SODA}, pages 291--300, 1993.

\bibitem{CG06}
H.~T.-H. Chan and A.~Gupta.
\newblock Small hop-diameter sparse spanners for doubling metrics.
\newblock In {\em Proc. of 17th SODA}, pages 70--78, 2006.

\bibitem{CGMZ05}
H.~T.-H. Chan, A.~Gupta, B.~M. Maggs, and S.~Zhou.
\newblock On hierarchical routing in doubling metrics.
\newblock In {\em Proc. of 16th SODA}, pages 762--771, 2005.

\bibitem{Chan08}
T.~M. Chan.
\newblock Well-separated pair decomposition in linear time?
\newblock {\em Inf. Process. Lett.}, 107(5):138--141, 2008.

\bibitem{Chandra94}
B.~Chandra.
\newblock Constructing sparse spanners for most graphs in higher dimensions.
\newblock {\em Inf. Process. Lett.}, 51(6):289--294, 1994.

\bibitem{Chaz87}
B.~Chazelle.
\newblock Computing on a free tree via complexity-preserving mappings.
\newblock {\em Algorithmica}, 2:337--361, 1987.

\bibitem{CR91}
B.~Chazelle and B.~Rosenberg.
\newblock The complexity of computing partial sums off-line.
\newblock {\em Int. J. Comput. Geom. Appl.}, 1:33--45, 1991.

\bibitem{Chew86}
L.~P. Chew.
\newblock There is a planar graph almost as good as the complete graph.
\newblock In {\em Proc. of 2nd SOCG}, pages 169--177, 1986.

\bibitem{DHN93}
G.~Das, P.~J. Heffernan, and G.~Narasimhan.
\newblock Optimally sparse spanners in 3-dimensional euclidean space.
\newblock In {\em Proc. of 9th SOCG}, pages 53--62, 1993.

\bibitem{DJ89}
G.~Das and D.~Joseph.
\newblock Which triangulations approximate the complete graph?
\newblock In {\em Proc. of the International Symp. on Optimal Algorithms,
  volume 401 of Lecture Notes in Computer Science}, pages 168--192, 1989.

\bibitem{DN94}
G.~Das and G.~Narasimhan.
\newblock A fast algorithm for constructing sparse {E}uclidean spanners.
\newblock In {\em Proc. of 10th SOCG}, pages 132--139, 1994.

\bibitem{DNS95}
G.~Das, G.~Narasimhan, and J.~S. Salowe.
\newblock A new way to weigh malnourished euclidean graphs.
\newblock In {\em Proc. of 6th SODA}, pages 215--222, 1995.

\bibitem{BCKO08}
M.~de~Berg, O.~Cheong, M.~van Kreveld, and M.~Overmars.
\newblock {\em Computational Geometry: Algorithms and Applications, third
  edition}.
\newblock Springer-Verlag, Heidelberg, 2008.

\bibitem{DES09tech}
Y.~Dinitz, M.~Elkin, and S.~Solomon.
\newblock Low-light trees, and tight lower bounds for {E}uclidean spanners.
\newblock {\em Discrete \& Computational Geometry}, 43(4):736--783, 2010.

\bibitem{FHP05}
J.~Fischer and S.~Har-Peled.
\newblock Dynamic well-separated pair decomposition made easy.
\newblock In {\em Proc. of 17th CCCG}, pages 235--238, 2005.

\bibitem{Fred93}
G.~N. Frederickson.
\newblock A data structure for dynamically maintaining rooted trees.
\newblock In {\em Proc. of 4th SODA}, pages 175--184, 1993.

\bibitem{GR08}
L.~Gottlieb and L.~Roditty.
\newblock An optimal dynamic spanner for doubling metric spaces.
\newblock In {\em Proc. of 16th ESA}, pages 478--489, 2008.

\bibitem{GLN02}
J.~Gudmundsson, C.~Levcopoulos, and G.~Narasimhan.
\newblock Fast greedy algorithms for constructing sparse geometric spanners.
\newblock {\em SIAM J. Comput.}, 31(5):1479--1500, 2002.

\bibitem{GLNS02}
J.~Gudmundsson, C.~Levcopoulos, G.~Narasimhan, and M.~H.~M. Smid.
\newblock Approximate distance oracles for geometric graphs.
\newblock In {\em Proc. of 13th SODA}, pages 828--837, 2002.

\bibitem{GLNS08}
J.~Gudmundsson, C.~Levcopoulos, G.~Narasimhan, and M.~H.~M. Smid.
\newblock Approximate distance oracles for geometric spanners.
\newblock {\em ACM Transactions on Algorithms}, 4(1), 2008.

\bibitem{GNS05}
J.~Gudmundsson, G.~Narasimhan, and M.~H.~M. Smid.
\newblock Fast pruning of geometric spanners.
\newblock In {\em Proc. of 22nd STACS}, pages 508--520, 2005.

\bibitem{HPM05}
S.~Har-Peled and M.~Mendel.
\newblock Fast construction of nets in low dimensional metrics, and their
  applications.
\newblock In {\em Proc. of 21st SOCG}, pages 150--158, 2005.

\bibitem{HP00}
Y.~Hassin and D.~Peleg.
\newblock Sparse communication networks and efficient routing in the plane.
\newblock In {\em Proc. of 19th PODC}, pages 41--50, 2000.

\bibitem{Keil88}
J.~M. Keil.
\newblock Approximating the complete euclidean graph.
\newblock In {\em Proc. of 1st SWAT}, pages 208--213, 1988.

\bibitem{KG92}
J.~M. Keil and C.~A. Gutwin.
\newblock Classes of graphs which approximate the complete euclidean graph.
\newblock {\em Discrete \& Computational Geometry}, 7:13--28, 1992.

\bibitem{LSW94}
H.~P. Lenhof, J.~S. Salowe, and D.~E. Wrege.
\newblock New methods to mix shortest-path and minimum spanning trees.
\newblock manuscript, 1994.

\bibitem{MP00}
Y.~Mansour and D.~Peleg.
\newblock An approximation algorithm for min-cost network design.
\newblock {\em DIMACS Series in Discr. Math and TCS}, 53:97--106, 2000.

\bibitem{NS07}
G.~Narasimhan and M.~Smid.
\newblock {\em Geometric Spanner Networks}.
\newblock Cambridge University Press, 2007.

\bibitem{PU89}
D.~Peleg and J.~D. Ullman.
\newblock An optimal synchronizer for the hypercube.
\newblock {\em SIAM J. Comput.}, 18(4):740--747, 1989.

\bibitem{PD04}
M.~{P\v{a}tra\c{s}cu} and E.~D. Demaine.
\newblock Tight bounds for the partial-sums problem.
\newblock In {\em Proc. of 15th SODA}, pages 20--29, 2004.

\bibitem{RS98}
S.~Rao and W.~D. Smith.
\newblock Approximating geometrical graphs via ``spanners'' and ``banyans''.
\newblock In {\em Proc. of 30th STOC}, pages 540--550, 1998.

\bibitem{RS91}
J.~Ruppert and R.~Seidel.
\newblock Approximating the $d$-dimensional complete {E}uclidean graph.
\newblock In {\em Proc. of 3rd CCCG}, pages 207--210, 1991.

\bibitem{Smid09}
M.~H.~M. Smid.
\newblock Private communication.

\bibitem{Sol11}
S.~Solomon.
\newblock An optimal-time construction of sparse euclidean spanners with tiny
  diameter.
\newblock In {\em Proc. of 22st SODA}, pages 820--839, 2011.

\bibitem{Tarj79}
R.~E. Tarjan.
\newblock Applications of path compression on balanced trees.
\newblock {\em J. ACM}, 26(4):690--715, 1979.

\bibitem{Thor92}
M.~Thorup.
\newblock On shortcutting digraphs.
\newblock In {\em Proc. of 18th WG}, pages 205--211, 1992.

\bibitem{Thor95}
M.~Thorup.
\newblock Shortcutting planar digraphs.
\newblock {\em Combinatorics, Probability \& Computing}, 4:287--315, 1995.

\bibitem{Thor97}
M.~Thorup.
\newblock Parallel shortcutting of rooted trees.
\newblock {\em J. Algorithms}, 23(1):139--159, 1997.

\bibitem{Yao82}
A.~C. Yao.
\newblock Space-time tradeoff for answering range queries.
\newblock In {\em Proc. of 14th STOC}, pages 128--136, 1982.

\end{thebibliography}
